\documentclass[12pt]{article}
\usepackage{amsmath,amsfonts}
\usepackage{hyperref}
\usepackage{graphicx}
\usepackage{xcolor}
\usepackage[numbers,sort&compress]{natbib}

\usepackage{amsthm}
\usepackage{amssymb}
\usepackage[matrix,arrow]{xy}
\usepackage{epsfig}
\usepackage{slashed}

\makeatletter\@addtoreset{equation}{section}\makeatother

\newcommand{\al}{\alpha}
\newcommand{\be}{\beta}
\newcommand{\te}{\theta}

\newcommand{\onov}[1]{\frac{1}{#1}}

\newcommand{\lag}{\mathcal{L}}

\newcommand{\beq}{\begin{equation}}
\newcommand{\eeq}{\end{equation}}
\newcommand{\bea}{\begin{eqnarray}}
\newcommand{\eea}{\end{eqnarray}}

\newcommand{\vev}[1]{{\left< {#1} \right>}}

\def\CT{\mathcal{T}}

\newcommand{\eq}[2][ ]{\begin{equation}\label{#1}{\begin{split}#2\end{split}}\end{equation}}
\newcommand{\eql}[2]{\begin{equation}\label{#1}{\begin{split}#2\end{split}}\end{equation}}

\newcommand{\Tr}{{\rm Tr\,}}
\newcommand{\tr}{{\rm tr\,}}

\newcommand{\cD}{{\mathcal D}}

\newcommand{\cN}{{\mathcal N}}

\newcommand{\cT}{{\mathcal T}}

\renewcommand{\title}[1]{\vbox{\center\LARGE{#1}}\vspace{5mm}}
\renewcommand{\author}[1]{\vbox{\center\large#1}\vspace{5mm}}
\newcommand{\address}[1]{\vbox{\center\em#1}}
\newcommand{\email}[1]{\vbox{\center\tt#1}\vspace{5mm}}

\begin{document}

\title{The Bi-Fundamental Gauge Theory in 3+1 Dimensions: The Vacuum Structure and a Cascade}
\author{Avner Karasik$^{1,\spadesuit}$ and Zohar Komargodski$^{1,2,\clubsuit}$}
\address{$^{1}$Weizmann Institute of Science, Rehovot 76100, Israel \\
$^{2}$ Simons Center for Geometry and Physics, Stony Brook, New York, USA}
\email{$^{\spadesuit}$avnerkar@gmail.com, $^{\clubsuit}$zkomargo@gmail.com}
 \thispagestyle{empty}
\abstract{We study the phases of the $SU(N_1)\times SU(N_2)$ gauge theory with a bi-fundamental fermion in 3+1 dimensions. We show that the discrete anomalies and Berry phases associated to the one-form symmetry of the theory allow for several topologically distinct phase diagrams. We identify several limits of the theory where the phase diagram can be determined using various controlled approximations. When the two ranks are equal $N_1=N_2$, these limits all lead to the same topology for the phase diagram and provide a consistent global understanding of the phases of the theory. When $N_1\neq N_2$, different limits lead to distinct topologies of the phase diagram. This necessarily implies non-trivial physics at some intermediate regimes of parameter space. In the large $N_{1,2}$ limit, we argue that the topological transitions are accounted for by a (non-supersymmetric) duality cascade as one varies the parameters of the theory.  }

\vspace{2em}
\today

\newpage

\bibliographystyle{utphys}
\tableofcontents
\section{Review and Summary}

Recently, new tools were introduced into the study of strongly coupled Quantum Field Theories.  In light of this, we would like to consider a particular class of strongly coupled four-dimensional gauge theories. We consider the gauge group to be the product group $SU(N_1)\times SU(N_2)$. The matter fields are given by a Weyl fermion $\psi$ in the representation $(\square,\bar \square)$ and another Weyl fermion $\tilde \psi$ in the representation $(\bar \square, \square)$. Together, $\psi$ and $\tilde \psi$ naturally combine into a Dirac fermion. The theory can be defined by the microscopic Lagrangian
\beq \label{YMin} {\cal L} = \sum_{k=1,2}\left[{1\over 4g_k^2}\Tr F_k^2+ i{\theta_k\over 8\pi^2} \Tr F_k\wedge F_k\right]+i\psi^\dagger \slashed{\cD}\psi+i\tilde{\psi}^\dagger\slashed{\cD}\tilde{\psi}+M \psi\tilde \psi+c.c.~.
\eeq  
The dimensionless parameters $g_i^2$ exist only classically; in the quantum theory they are replaced by the dimensionful scales $\Lambda_i$, at which the gauge inetractions become strong. $\theta_i$ are dimensionless parameters in the full quantum theory and they are valued in the circle $\theta_i\simeq \theta_i+2\pi$. $M$ can be taken without loss of generality to be real and non-negative such that the two theta angles $\theta_1,\theta_2$ are physical. 
The infrared behaviour of the theory therefore depends on the non-negative number $M$ 
(in units of one of the strong coupling scales), the ratio $\Lambda_1/\Lambda_2$, and the two circle-valued theta angles $\theta_1,\theta_2$.

This theory is tractable in several corners of the parameter space. The large mass limit $M\gg \Lambda_i$ can be studied by integrating out the fermion and computing corrections to the two decoupled $SU(N_i)$ Yang-Mills (YM) theories. The small mass limit $M\ll \Lambda_i$ is also often tractable. This is because in the massless limit $M=0$, the classical Lagrangian is invariant under chiral rotations of the fermion. Quantum mechanically, depending on $N_1,N_2$ a discrete chiral symmetry survives and this symmetry constrains the dynamics. 

Another tractable corner is the hierarchical limit $\Lambda_1\gg\Lambda_2$. In this limit we can use chiral Lagrangian techniques to study the phase diagram of the theory. Finally, the large $N_1,N_2$ limit is tractable since in that limit the dependence on the theta angles is subleading (and locally quadratic).

Intermediate regimes of the parameter space are more difficult to study and this is where we appeal to some recent developments in the study of strongly coupled theories. One can identify certain Renormalization-Group (RG) invariants of the theory which allow to make rigorous statements about the infrared for arbitrary values of the couplings.
These quantities, which are inert under the RG, are related to 't Hooft anomalies and certain generalizations which we will discuss in detail (also sometimes referred to in the literature as ``global inconsistency conditions'').

While the concept of 't Hooft anomalies is very familiar, the concept of ``global inconsistency''~\cite{Gaiotto:2014kfa, Gaiotto:2017yup,Komargodski:2017dmc,Thorngren:2017vzn,VideoRyan} is perhaps less familiar. In fact, ``global inconsistency'' is simply a certain generalization of the Berry phase. Consider a family of theories labeled by a periodic parameter, $\Theta\in [0,2\pi]$ (this could be any periodic parameter, including the $\theta$ parameter in Yang-Mills theories~\eqref{YMin}). The partition function is a function of this parameter. However, upon coupling the theory to background fields for various global symmetries (which could be zero-form symmetries or higher-form symmetries~\citep{Kapustin:2014gua,Gaiotto:2014kfa}) the partition function may no longer be a function of $\Theta$ but a section of some bundle. Let us denote these background fields collectively by $A$. Then the partition function satisfies \begin{equation}\label{BP}Z[\Theta+2\pi]=Z[\Theta]e^{i\mathcal{S}[A]}~,\end{equation}
where $\mathcal{S}[A]$ is a local action for $A$ (in general, a local action is an invertible field theory).  Of course $\mathcal{S}[A]$ exists only in a cohomological sense -- we can always modify the original theory by a local counter-term that depends on $\Theta$ and may shift $\mathcal{S}[A]$ as a result. 
However, if $\mathcal{S}[A]$  cannot be removed by such a redefinition of the theory, it has interesting implications for the ground state of the theory. Indeed, suppose the ground state of the theory is trivial and gapped for all $\Theta$. This leads to an immediate contradiction because under these assumptions, for all $\Theta$, the partition function is given by some integral of a local object $\hat{\mathcal{S}}[A,
\Theta]$ (we can throw away all the pieces which are suppressed by the gap) and such local terms cannot, by assumption, lead to the Berry phase~\eqref{BP} for cohomological reasons.\footnote{It turns out one can think of the global inconsistency as an anomaly if one includes background fields for a -1-form symmetry~\cite{Upcoming}.}

Let us review some of these concepts in the particular case of pure $SU(N)$ Yang-Mills theory. There are two distinguished values of $\theta$ where the theory has time reversal symmetry (alternatively, parity): These are $\theta=0$ and $\theta=\pi$. The discussion is slightly different for odd and even $N$.

\begin{itemize}
	\item {\underline{even $N$:}} For $\theta=\pi$ the ground state of the theory cannot be trivial. Indeed, there exists a 't Hooft anomaly involving time reversal symmetry and the one-form $\mathbb{Z}_N$ symmetry associated to the center of the group $SU(N)$. This anomaly precludes a trivial non-degenerate confining vacuum at $\theta=\pi$. By contrast, at $\theta=0$ the theory is widely believed to have a trivial non-degenerate confining vacuum. 
	The properties of the ground state at $\theta=\pi$ are strongly constrained by the above-mentioned anomaly. One option is that the theory confines but time reversal symmetry is spontaneously broken and then the anomaly is saturated by the nontrivial physics on the domain wall between the two degenerate vacua. This is certainly what happens for large enough $N$~\cite{Witten:1980sp,Witten:1998uka}.\footnote{For small values of $N$, the infrared theory could be different (as long as it is consistent with the anomaly). In particular, it could be that the theory at $\theta=\pi$ is deconfined. Note that for $\theta=\pi$ the theory could be put on a non-orientable manifold and this may lead to additional constraints on the infrared dynamics. See recent discussions in~\cite{Wan:2019oyr} and references therein.} 
	\item {\underline{odd $N$:}} Here the situation is a little more complicated. One encounters the phenomenon of global inconsistency explained above (we will review this in detail in the main body of the text). What it means is that the theory at long distances cannot be trivial for all values of $\theta$. There must be nontrivial infrared physics at least at one value of $\theta$. In practice, the theory is widely believed to have a trivial confined ground state away from $\theta=\pi$ (this is certainly true at large enough $N$) and hence the consequences of this global inconsistency are in this particular case similar to the discussion above for even $N$. 
	
\end{itemize}

Returning to our bi-fundamental gauge theory, the constraints of anomalies and global inconsistency conditions alone turn out to be compatible with several topologically distinct phase diagrams. A very closely related study of the anomalies was carried out in~\cite{Tanizaki:2017bam} (here we discuss a different question, regarding the phases of the theory in infinite space). 

Most of our detailed analysis is for the case where the two ranks are equal $N_1=N_2$. We are able to determine the topology of the phase diagram in several limits, such as the limits of small and large $M$ as well as $\Lambda_1/\Lambda_2 \gg1 $ and $N_1=N_2\to\infty$. It is encouraging that the different limits lead to the very same topology for the phase diagram. Furthermore, our results in the different limits are consistent with the anomalies and global inconsistency conditions. This is encouraging because analyzing these different limits requires entirely different tools and ideas in field theory. 

\begin{figure}
	\vspace{10pt}
	\begin{center}
		\includegraphics[scale=0.3]{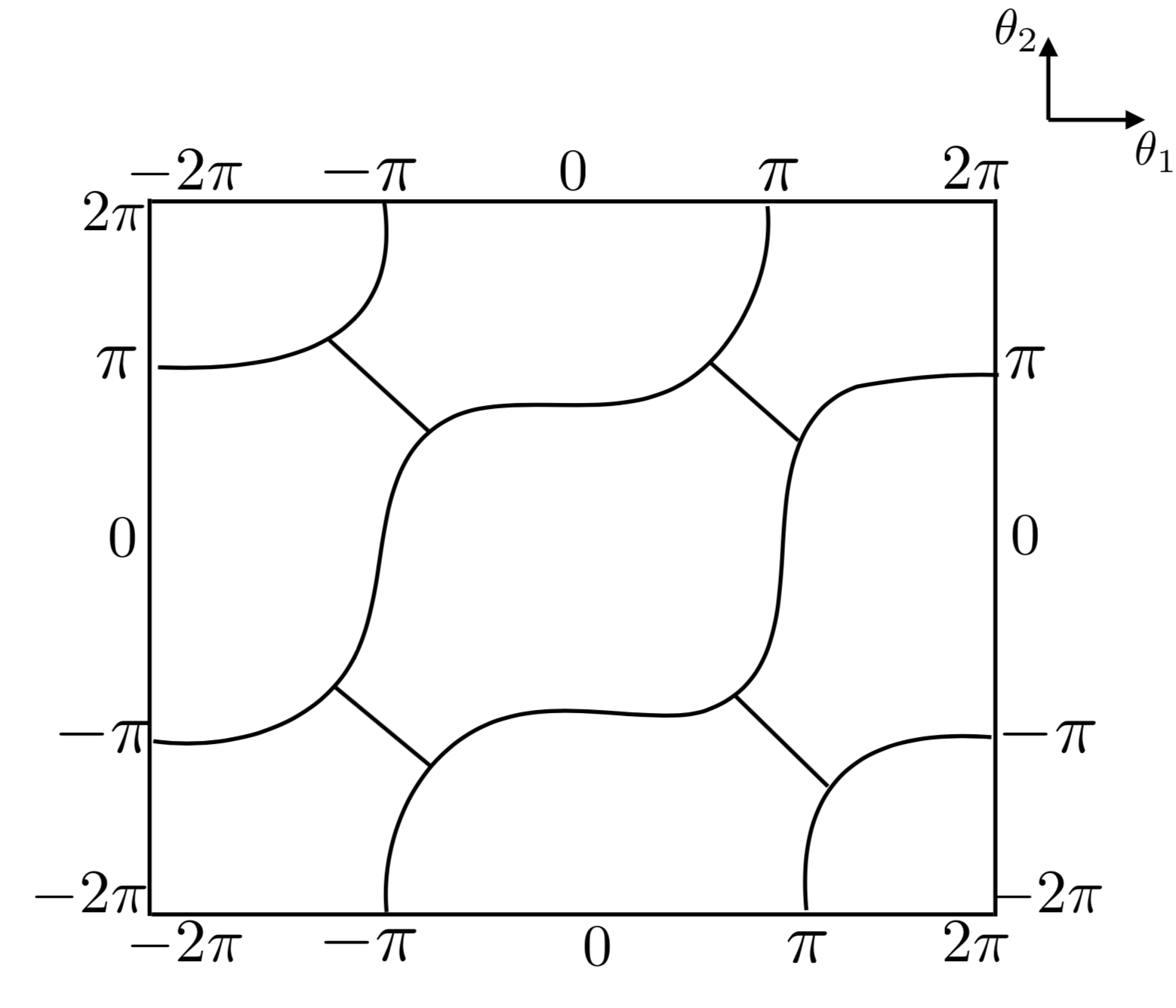}
		\caption{\small{ The topology of the phase diagram of the equal rank bi-fundamental gauge theory for some fixed $M,\Lambda_1,\Lambda_2$ as a function of $\theta_{1,2}$.}}\label{topology}
	\end{center}
	\vspace{10pt}
\end{figure}

Hence we are able to arrive at a uniform, consistent picture for the dynamics of the equal-rank bi-fundamental gauge theory. The proposal for the equal-rank case is that the topology of the phase diagram in the $(\te_1,\te_2)$ plane is the same for every finite $M,\ \Lambda_1,\ \Lambda_2$. Changing $M,\ \Lambda_1,\ \Lambda_2$ with respect to each other modifies the diagram smoothly without changing its topology. The topology of the phase diagram is presented in figure~\ref{topology}.

The curves in~ figure \ref{topology} stand for a doubly degenerate ground state (when the curves intersect in a trivalent vertex, the degeneracy is threefold). This double degeneracy is generally {\it not} due to spontaneous breaking of some global symmetry! The degeneracy is forced to exist by the Berry phase and anomalies that we explain in the main body of the text. 

\begin{figure}
	\vspace{10pt}
	\begin{center}
		\includegraphics[scale=0.3]{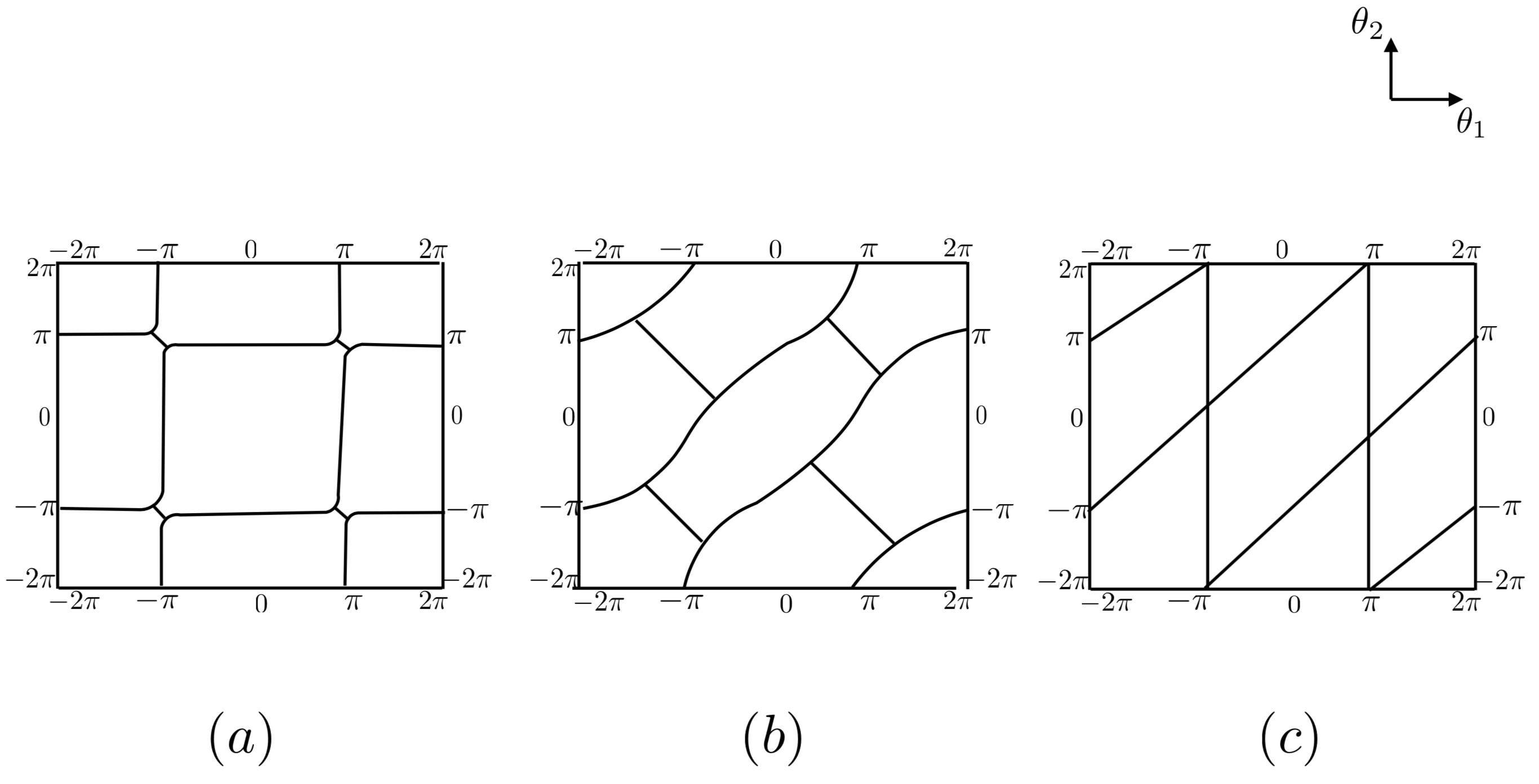}
		\caption{\small{The qualitative structure of the phase diagram of the bi-fundamental gauge theory with $N_1=N_2$ in three limits: (a) Large $M$, (b) Small $M$ with $\Lambda_1\simeq \Lambda_2$, (c) $\Lambda_2, M\ll\Lambda_1$. }}\label{cases}
	\end{center}
	\vspace{10pt}
\end{figure}

In various limits of the theory one can compute the shape of the phase transition lines. Let us review the main three limits we study, which are respectively depicted in figure~\ref{cases} (we are still restricting ourselves to the equal-rank case).

\begin{itemize}

\item If $M\gg \Lambda_{1,2}$, we can approximate the dynamics by the dynamics of two pure $SU(N)$ gauge theories coupled by irrelevant operators. To determine those irrelevant operators, we carry out an explicit one-loop computation. If not for these irrelevant operators, the phase diagram would look like a cross and at the quadrivalent vertex there would be four degenerate vacua. This quadrivalent vertex has two topologically inequivalent resolutions which are both consistent with the anomalies and Berry phases. The computation of the irrelevant operator is necessary to determine which of the resolutions takes place. 

\item If $M\ll \Lambda_{1,2}$ we can analyze the theory in an expansion around the massless theory, which has a discrete chiral symmetry and a single physical theta angle. The mass lifts slightly the degeneracy of the massless theory and this leads to a phase diagram similar to that in figure~\ref{cases} (b), where all the lines are approximately straight. 

\item For $\Lambda_2\ll M\ll \Lambda_1$, the idea is that while the first group becomes strong, the mass and the dynamics of the second group are still negligible. This allows to convert the problem to gauging a symmetry on the $SU(N_2)$ group manifold, with some mass. This is still a hard problem, though in the particular limit we define in the text (where the mass $M$ is small also compared to some radiative effects in $g_2$), one can say a few concrete  things about the problem. This leads to the diagram of figure~\ref{cases}(c).
Note that the phase diagram in figure~\ref{cases}(c) still has quadrivalent vertices. Indeed, to resolve that singularity one has to compute yet another (small) radiative correction which we have not done. Yet, the effects we have been able to account for in this limit lead to some new consistency checks.

\end{itemize}

In addition to these various simplifying limits, we also study the 't Hooft limit, where the dependence on $(\theta_1,\theta_2)$ is severely constrained to be a locally quadratic function. We show that the phases we propose are consistent with the 't Hooft limit.

\begin{figure}
	\vspace{10pt}
         	\includegraphics[width=0.9\textwidth, height=6.5cm]{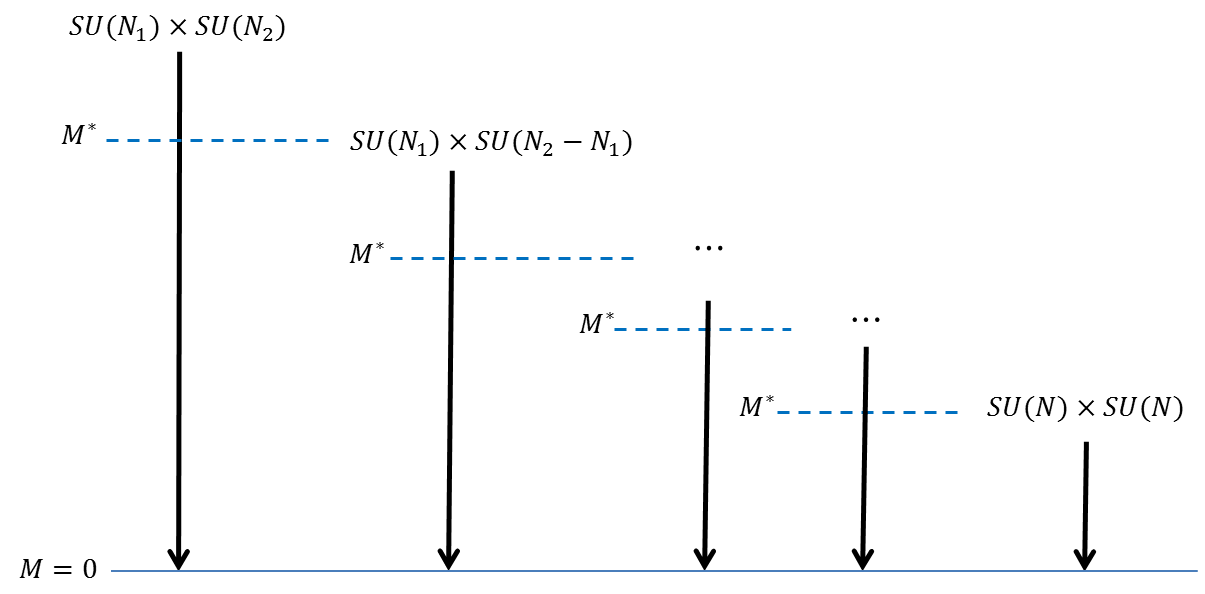}
  \caption{\small{Schematic illustration of the duality cascade. Starting from $SU(N_1)\times SU(N_2)$ on the left, as we lower the mass of the fermion we uncover dual theories that describe the correct evolution of the phase diagram. The last step of the cascade is the $SU(N)\times SU(N)$ theory where $N=gcd(N_1,N_2)$ (assuming here $N\gg1$). The massless point $M=0$ is the same for all the theories.}}\label{cascade}
	\vspace{10pt}
\end{figure}

We also explore of the more general case, where the two ranks are different $N_1\neq N_2$. The main novelty here is that the different calculable limits no longer lead to the same topology of the phase diagram. This means that as the parameters are varied, the topology of the diagram changes. There are special surfaces in the parameter space $M=M^*(\Lambda_1,\Lambda_2)$  where the topology transition takes place. At these loci there are fourfold vacuum degeneracies for some $\theta_1,\theta_2$. A curious observation is that in the large $N_{1,2}$ limit, the topology transition around $M=M^*$ is captured by $SU(N_1)\times SU(N_2-N_1)$ gauge theory with a bi-fundamental Dirac fermion (assuming here $N_2>N_1$ without loss of generality). This leads to a structure of IR dualities similar to the supersymmetric duality cascade found by Klebanov and Strassler in \cite{Klebanov:2000hb}. This is illustrated in figure~\ref{cascade}.

Let us now make some brief comments about various other questions to which the present study is related.  
\begin{itemize}
\item {\underline{Domain Walls:}} Considering the phase diagram in figure~\ref{topology} we see that there are various situations with two or even three degenerate ground states. 
We can then consider domain wall configurations interpolating between these ground states. While the physics of the bulk vacua may be continuous (for equal rank) as we dial the parameters of the microscopic theory, the domain wall could undergo phase transitions in the universality class of certain 2+1 dimensional gauge theories. More concretely, we expect the domain wall theories to be in the universality class of the bi-fundamental 2+1 dimensional gauge Chern-Simons theory (related models were recently studied in~\cite{Aitken:2018cvh}). In the context of the observed duality cascade for different ranks, it would be very interesting to see the same structure also from the domain wall point of view. (Note again that the phase diagram allows for several inequivalent types of domain walls.)
Similar questions were recently studied in other examples, see for instance \cite{Komargodski:2017smk, Gaiotto:2017tne, Draper:2018mpj, Bashmakov:2018wts, Ritz:2018mce, Choi:2018tuh, Bashmakov:2018ghn,Rocek:2019eve,Argurio:2018uup}.  

\item {\underline{Duality cascade:}} The duality cascade of \cite{Klebanov:2000hb} concerns with $\cN=1$ supersymmetric bi-fundamental theories and has a very similar structure to the one we found in the non-supersymmetric case. An interesting difference between the two is that while we work in the IR and encounter the cascade by changing the mass of the fermion, the original cascade arises by changing the RG scale. It would be very interesting to understand better the relation between the two cascades and whether this structure exists also for other similar theories.

\item {\underline{Planar Equivalence:}} The bi-fundamental gauge theory arises naturally in the context of planar equivalence and volume independence. For discussions involving this particular theory see for instance~\cite{Strassler:2001fs,Gorsky:2002wt,Tong:2002vp,Kovtun:2004bz,Armoni:2005wta}. Our proposal for the phase diagram in the various limits can be further studied by the circle reduction techniques of
~\cite{Unsal:2007jx, Poppitz:2012sw, Poppitz:2012nz}. It would be interesting to see if the results match. 

\item {\underline{String Constructions:}} The bi-fundamental gauge theory is closely related to the work of~\cite{Majumder:1999yy}.\footnote{We thank C. Vafa for bringing this to our attention.} There are also other ways to probe the dynamics of this theory via brane constructions. 

\end{itemize} 

The outline of our note is as follows. In section~\ref{sunym} we review how the Berry phases and anomalies constrain the dynamics of pure Yang-Mills theory with gauge group $SU(N)$. Then we move on to analyze the equal rank bi-fundamental gauge theory. In section \ref{bifundamental} we compute the Berry phases and anomalies of the theory and study the large mass limit. In section~\ref{massless} we study the massless theory. The massless theory still has a physical theta angle and displays some novel phenomena as a result. In section~\ref{smallm} we include small mass corrections. In section~\ref{chirallag} we study the physics in the limit of $\Lambda_2\ll M\ll \Lambda_1$. In section~\ref{diffrank} we move on to the case where the ranks of the two gauge groups differ and encounter the cascade phenomenon. We finish with some remarks and speculations. Appendix~\ref{1loop} covers the details of a certain essential one-loop computation.

\section{$SU(N)$ Yang-Mills Theory}
\label{sunym}
In this section we will summarize some of the results of \cite{Gaiotto:2017yup} regarding the phase diagram of pure $SU(N)$ Yang-Mills (YM) in four dimensions. The theory is parametrized by the $\te$ angle and the gauge coupling $g$ that transmutes to the strong coupling scale $\Lambda$. 
The theory has a $\mathbb{Z}_N$ one-form symmetry associated to the center of the gauge group. This is a global one-form symmetry \citep{Kapustin:2014gua,Gaiotto:2014kfa}. The charged objects under this symmetry are the line operators of the theory. Consider the Wilson line in the fundamental representation $W_F=\tr_F Pe^{i\int A}$, then under the action of the generator of the $\mathbb{Z}_N$ symmetry, it transforms as $W_F\to e^{2\pi i/N}W_F$. We say that this one-form symmetry is unbroken if the theory is confined and the Wilson line obeys an area law. 

Another symmetry that arises at $\te=0$ and $\te=\pi$ is time reversal symmetry, denoted by $\CT$. $\CT$ takes $\te\to-\te$. For $\te=0$ this is obviously a symmetry. For $\te=\pi$ this is a symmetry due to the $2\pi$ periodicity of $\te$, which means that $\te=\pi$ and $\te=-\pi$ obey the same physics. However, symmetries that involve shifting of $\te$ by integer multiple of $2\pi$ require non-trivial identification of the line operators~\cite{Witten:1979ey} and will often have mixed anomalies with the one-form symmetry. In order to see it explicitly, we can couple the theory to a $\mathbb{Z}_N$ discrete two-form gauge field, $B$, which is a background gauge field for the one-form symmetry. 

The main point is that in the presence of such a background two-form gauge field, the partition function is no longer periodic under $\theta\to\theta+2\pi$. The non-periodicity is due to a c-number, as follows 
\begin{equation}\label{tranrule}Z_{YM}[\theta+2\pi]=Z_{YM}[\theta]e^{{2
\pi i(N-1)\over 2N}\int_{\mathcal{M}_4} B\cup B}~, \end{equation}
where $\mathcal{M}_4$ is the space-time four-manifold and $B\cup B$ is an imprecise notation -- it stands for the Pontryagin square.

In preparation for the analysis of anomalies and global inconsistency conditions, let us now review the possible counter-terms. The analysis of the allowed counter-terms below is done only for spin manifolds (partly in order not to complicate the discussion and partly because the theories of interest in this paper have fermions). 
The most general allowed ultraviolet counter-term is of the form 
\begin{equation}\label{counter}\frac{2\pi i p}{2N}\int_{\mathcal{M}_4} B\cup B~,\end{equation} 
where for 
\begin{itemize}
\item even $N$: $p$ is allowed to take any integer value. However only its value mod $N$ matters.
The counterterm~\eqref{counter} is unchanged if we shift $p$ by any multiple of $N$. This is not obvious. It is has to do with the fact that for spin manifolds $B\cup B$ is an even form. In other words, on spin manifolds $e^{\pi i \int_{\mathcal{M}_4} B\cup B}=1$.
 
\item odd $N$: $p$ is allowed to take only even values. Furthermore, if we shift $p$ by an integer multiple of $2N$ the counter-term is unchanged. 
\end{itemize}

We see that the space of allowed counter-terms is somewhat different for odd and even $N$ and this would lead to somewhat different physical conclusions. First of all, since $p$ is quantized, the counter-term cannot depend on $\theta$ and hence the Berry phase in~\eqref{tranrule} is physical. This already implies that for {\it some} value of $\theta$ the ground state cannot be trivial. We can try to further ask if there is an 't Hooft anomaly at either $\theta=0$ or $\theta=\pi$. 

Since time reversal symmetry at $\te=\pi$ is accompanied by a shift of $\theta$ by $2\pi$, the phase in~\eqref{tranrule} would imply that, in the presence of the background field $B$, the time reversal symmetry at $\theta=\pi$ is explicitly broken. However, as always, we should allow for a counter-term in the ultraviolet and attempt to restore the time reversal symmetry at $\te=\pi$. 
For both even and odd $N$, it is clear that time reversal symmetry takes $p\to-p$. 

Let us therefore add in the ultraviolet a counter-term with some (allowed) $p$. First, using~\eqref{tranrule} we see that under $\theta\to \theta+2\pi$ we get
$${\rm odd} \ N \ : \ p\to p+N-1~.$$
This formula makes sense for odd $N$ since if $p$ is even then so is  $p+N-1$. For even $N$ we can simplify the transformation since on spin manifolds $p$ is only defined mod $N$ and hence we can write for even $N$
$${\rm even} \ N 
 \ : \
 p\to p-1~.$$

Now we are ready to examine the question if there is a conflict between coupling the theory to a background field $B$ and time reversal symmetry.

For $\te=0$, we can always choose the counter term with $p=0$. Therefore, there is never a conflict between time reversal symmetry and the one-form $\mathbb{Z}_N$ symmetry. But this is not the only allowed choice of the counter-term at $\te=0$. For even $N$, we could choose $p$ such that $2p=0 \ {\rm mod} \ N$, i.e. also $p=N/2$ works. For odd $N$ the general ultraviolet counter-term consistent with time reversal symmetry would satisfy $2p=0 \ {\rm mod} \ 2N$. Remembering that $p$ must be even, and that it is defined mod $2N$, it is clear that there are no solutions other than $p=0$ in this case.

 Now let us discuss $\te=\pi$. Since we need to accompany $\CT$ with the shift $\te\to\te+2\pi$ we are led to the following transformation of $p$ for even $N$:
$$\CT\ {\rm at} \ \te=\pi \ :  \ p\to -p-1 $$
and for odd $N$:
$$\CT\ {\rm at} \ \te=\pi \ :  \ p\to -p+N-1\ .$$

For even $N$, to retain the time reversal symmetry we therefore need to solve the equation $2p+1=0 \ {\rm mod} \ N$. Clearly, this has no solutions. In this situation we say that at $\te=\pi$ in the presence of the two-form gauge field $B$ it is impossible to retain time reversal symmetry, i.e. there is a 't Hooft anomaly involving the time reversal symmetry and the one-form $\mathbb{Z}_N$ symmetry. 
The anomaly leads to a further interesting constraint on the IR physics: the ground state cannot be trivial at $\theta=\pi$. 
For instance, if the IR theory is confining and trivially gapped, then time reversal symmetry must be spontaneously broken at $\theta=\pi$.

For odd $N$ we need to solve the equation $2p-N+1=0 \ {\rm mod} \ 2N$, for even $p$. For $N=1\ {\rm mod} \ 4$ the solution is $p={N-1\over 2}$ and for  $N=3\ {\rm mod} \ 4$ the solution is 
$p={N-1\over 2}+N$. Either way, it appears that for odd $N$ there is no conflict between time reversal symmetry and the one-form symmetry.  There is no 't Hooft anomaly in the traditional sense. However, the required counterterm is different at $\te=0$ and at $\te=\pi$. There is no one choice of counterterm that preserves $\CT$ at $\te=0$ and at $\te=\pi$ {\it simultaneously.} This is equivalent to the statement that there is global inconsistency. Indeed, the Berry phase in ~\eqref{BP} is non-trivial for all $N$.

Of course in the particular case of pure Yang-Mills theory with gauge group $SU(N)$, we do expect that odd $N$ and even $N$ should not differ much, at least for sufficiently large $N$ and hence in practice the global inconsistency would result in essentially the same fact, i.e. that there is a nontrivial ground state for $\theta=\pi$.\footnote{Such discrete anomalies and global inconsistency conditions also appear in lower-dimensional theories and can be used to place interesting constraints on the dynamics of strongly coupled theories. For some recent work on the subject see also~\cite{Benini:2017dus,Yamazaki:2017ulc, Gomis:2017ixy, Tanizaki:2017qhf, Tanizaki:2018xto, Ohmori:2018qza}. In addition, four-dimensional gauge theories may enjoy various other discrete anomalies, not of the type we have investigated here. For some recent related literature see~\cite{Anber:2018tcj,Cordova:2018acb,Bi:2018xvr,Wang:2018qoy,Wan:2018djl}.}

In the next section we will study in detail the theory with gauge group $SU(N)\times SU(N)$ and with a  Dirac fermion in the bi-fundamental representation. We will first study the constraints from anomalies and global inconsistency conditions and then proceed to some aspects involving the dynamics in various limits.

\section{The Bi-fundamental Gauge Theory}
\label{bifundamental}
In this section we embark on our study  of the phase diagram of four-dimensional $SU(N)\times SU(N)$ gauge theory with a bi-fundamental Dirac fermion. 

The classical Lagrangian is parametrized by the two gauge couplings $g_{1,2}$, the two periodic $\te$ angles $\te_{1,2}$ and the mass of the fermion $M$. Without loss of generality, we take the mass $M$ to be real and positive, and allow for general values for the two $\te$ angles. This theory has the following symmetries (we distinguish $M=0$ from $M\neq 0$):

\begin{center}$$\underline{M\neq 0}$$\end{center}
\begin{itemize}
\item $\mathbb{Z}_N$ one form symmetry, which is the diagonal subgroup of the center of the group $SU(N)\times SU(N)$: Indeed, under the center of $SU(N)\times SU(N)$ the fermion transform as $\psi\to e^{2\pi i k_1/N}\psi e^{-2\pi i k_2/N}$. The fermion is invariant if $k_1=k_2$. This diagonal $\mathbb{Z}_N$ is a global symmetry that acts only on line operators \cite{Gaiotto:2014kfa}.
Notice that in the $M\to\infty$ limit, the fermion decouples and the $\mathbb{Z}_N\times \mathbb{Z}_N$ one-form symmetry is restored.
\item $\CT$-Time reversal symmetry: $\CT$ takes $\te_i\to-\te_i$. Due to the $2\pi$ periodicity of the $\te$-angles, $\CT$ is a symmetry at all the following four choices for $(\te_1,\te_2)$: $(0,0),(0,\pi),(\pi,0),(\pi,\pi)$ (and all the images by $2\pi$ shifts).
\item $\mathcal{E}$- exchange symmetry: This is a $\mathbb{Z}_2$ transformation that exchanges the two gauge groups. It is a symmetry for $g_1=g_2$ and $\te_1=\te_2$ (mod $2\pi$).
Similarly, for $g_1=g_2$ and $\te_1=-\te_2$ (mod $2\pi$), the combined transformation $\mathcal{E}\times \CT$ is a symmetry. 
\item $U(1)_V$: Rotations of the fermion by a phase $\psi\to e^{i\al}\psi$, $\tilde \psi\to e^{-i\al}\tilde\psi$.

\begin{center}$$\underline{M=0}$$\end{center}

\item If the fermion is massless, the classical Lagrangian is invariant under chiral transformations $\psi\to e^{i\omega}\psi$, $\tilde \psi\to e^{i\omega}\tilde\psi$. However, due to the chiral (ABJ) anomaly this symmetry is broken to the subgroup $\mathbb{Z}_{2N}$ generated by $\omega=\frac{2\pi k}{2N}$ with integer $k$. The case $k=N$ is nothing but fermion number $(-1)^F$ (for even $N$ this is a gauge symmetry so the axial symmetry is reduced to $\mathbb{Z}_{N}$).

Due to the same chiral anomaly, one of the theta-angles is not physical in the massless theory. Indeed, we can set without loss of generality $\theta_1+\theta_2=0$. The physics of this massless model  depends on the linear combination $\te_-=\te_1-\te_2$ with $\te_-\simeq\te_-+2\pi$.
\end{itemize}

\subsection{Some Discrete Anomalies of the Bi-Fundamental Theory}


To analyze some of the anomalies and global inconsistency conditions for $M\neq 0$ we again study the effects of shifting the various theta angles by $2\pi$. 
As in the previous section, we couple the theory to a $\mathbb{Z}_N$ two-form gauge field, $B$, which acts as the background gauge field for the center symmetry~\cite{Kapustin:2014gua}. And as before we allow for a general counter-term, $$\frac{2\pi i p}{2N}\int B\cup B$$ where $p$ is an integer restricted as was explained after~\eqref{counter}. Shifting either $\theta_1$ or $\theta_2$ by $2\pi$ leads to the transformation rule 
$${\rm odd} \ N \ : \ p\to p+N-1~,$$
$${\rm even} \ N 
 \ : \
 p\to p-1~,$$
as before. In addition, time reversal symmetry simply reverses $p$. With these rules we can now examine the four time reversal invariant points $(0,0),(0,\pi),(\pi,0),(\pi,\pi)$. We examine the anomalies in detail for the cases of even $N$ and odd $N$ seperately. 

\newpage

\begin{center}  $$\underline{ {\rm even} \ N}$$\end{center}

\begin{itemize}
\item $(0,0)$: Here we need to solve $2p=0  \ {\rm mod} \ N$. This can be satisfied with either $p=0 $ or $p=N/2 \ {\rm mod} \ N$.
\item $(0,\pi)$ or $(\pi,0)$: we need to solve $2p+1=0  \ {\rm mod} \ N$, which clearly has no solutions.
\item $(\pi,\pi)$: we need to solve $2p+2=0  \ {\rm mod} \ N$. There exist the following solutions: $p=-1$ and $p=-1+{N/2}$.

\end{itemize}

We see that at $(0,\pi)$ and $(\pi,0)$ there is a bona fide 't Hooft anomaly involving time reversal symmetry and the one-form symmetry. The ground state at those points thus cannot be trivial. For instance, the anomaly at those points may be saturated by breaking time reversal symmetry spontaneously. 

The situation at $(\pi,\pi)$ is more involved. There is no bona fide 't Hooft anomaly since there exist counter-terms that allow to preserve all the symmetries. But those counter-terms do not coincide with those at $(0,0)$ and hence one encounters a global inconsistency type of situation.\footnote{$N=2$ is a notable exception (it is the only exception to what is said in the main text). Note that the counter-terms for $(\te_1,\te_2)=(0,0)$ and $(\te_1,\te_2)=(\pi,\pi)$ in that case precisely coincide and there is no global inconsistency between the two points. In the main text we assume $N>2$.} 

Equivalently, we see that regardless of the counter-terms that we are allowed to add, 
\begin{equation}\label{tranrulebi}Z_{YM}[\theta_1+2\pi,\theta_2+2\pi]=Z_{YM}[\theta_1,\theta_2]e^{{-2
\pi i\over N}\int_{\mathcal{M}_4} B\cup B}~, \end{equation}
which is a nontrivial Berry phase for all even $N>2$. 
The implication of this global inconsistency is that, if one draws any continuous trajectory connecting $(0,0)$ and $(2\pi,2\pi)$, one must encounter at least one point on the way which has a nontrivial ground state. We will see that this has far reaching consequences for the allowed phase diagrams for this model. 
\begin{center}  $$\underline{ {\rm odd} \ N}$$\end{center}
 Repeating the same analysis for odd $N$, we find that for a generic point of the form $(\te_1,\te_2)=(\pi k_1,\pi k_2)$, $\cT$ can be preserved by a counterterm satisfying
\eq{p=-p+(N-1)(k_1+k_2)\ \mod\ 2N\ ,}
where now $p$ is an even integer. This equation can always be solved by choosing $$p=\frac{(N-1)(k_1+k_2)}{2}+Nk\ ,\ k=0\ \text{or}\ 1\ .$$ It is easy to check that for every $N,\ k_1,\ k_2$, one of the two options gives an even $p$. We see that for odd $N$ we never encounter anomalies, but there are still global inconsistencies. In particular, we find the following Berry phase
\eq{Z_{YM}[\te_1+2\pi,\te_2+2\pi]=Z_{YM}[\te_1,\te_2]e^{-\frac{2\pi i}{N}\int_{\mathcal{M}_4}B\cup B}\ ,}
which is non-trivial for every odd $N>1$. Similar to the case of even $N$, also here the implication is that any continuous trajectory connecting $(0,0)$ and $(2\pi,2\pi)$ must encounter at least one point with a non-trivial ground state.

\subsection{The Decoupling Limit $M=\infty$}
\begin{figure}
	\vspace{10pt}
	\begin{center}
		\includegraphics[width=1\textwidth, height=10cm]{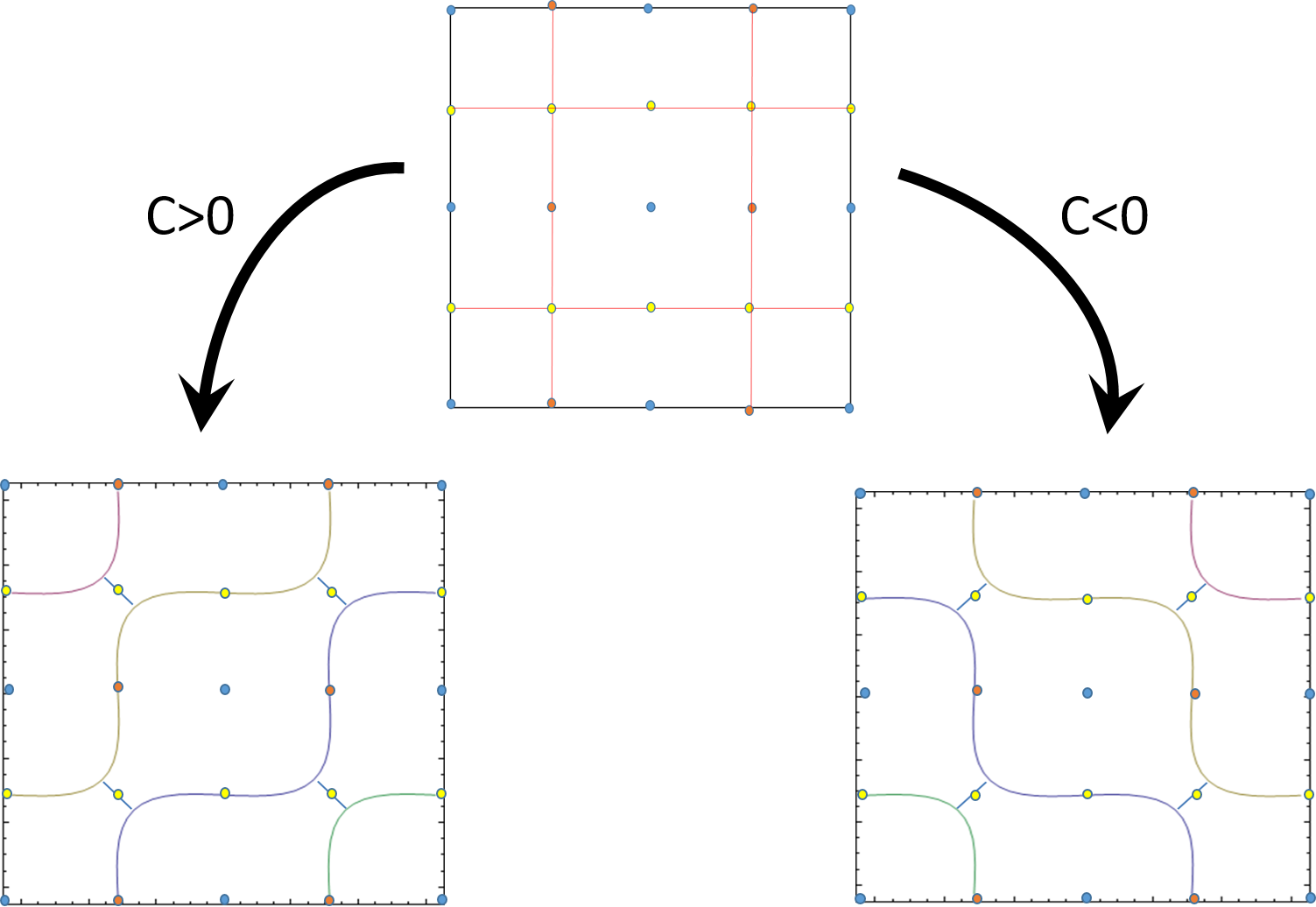}
		\caption{\small{The upper figure represnts the phase diagram of two decoupled $SU(N)$ theories. The leading correction in $\Lambda/M$ that comes from \eqref{leadingop} leads to two possible phase diagrams depending on the sign of $c$.  In these figures we include four copies of the fundamental domain.}}\label{2options}
	\end{center}
	\vspace{10pt}
\end{figure}

When the mass of the fermion is infinite, we get two decoupled $SU(N)$ gauge theories. 
For every one of those theories, there is a spontaneous symmetry breaking of $\CT$ at $\te=\pi$ (this is true for sufficiently large, finite, $N$, which we assume throughout). The phase diagram looks like the upper picture in figure \ref{2options}. On the lines, the vacuum is twofold degenerate and at the crossing of the two lines, it is fourfold degenerate. The vacuum is fourfold degenerate when both $\te_{1,2}=\pi$. 
The order parameters for the breaking of time reversal symmetry are the condensates 
\begin{equation}\label{fourgs}\langle \star TrF_{1}\wedge F_{1}\rangle\sim \pm\Lambda_{1}^4~,\qquad \langle \star TrF_{2}\wedge F_{2}\rangle\sim \pm \Lambda_2^4~.\end{equation} Since the sectors are decoupled there is time reversal symmetry separately for each of the $SU(N)$ gauge theories and at $\theta_{1,2}=\pi$ this leads to a fourfold degenerate ground state. 

Now let us take the fermion mass $M$ to be large (positive) and finite. The fourfold degeneracy of the vacuum is not a generic singularity and should be resolved in some way. Indeed, now time reversal symmetry must act on the two sectors together and the fourfold degeneracy should be lifted. Considering $\theta_{1,2}=\pi$, which is invariant under time reversal symmetry, the four ground states~\eqref{fourgs}, furnish a reducible representation of time reversal symmetry. Denoting the four ground states by $(+,+),(+,-),(-,+),(-,-)$ corresponding to the four possible sign combinations for the condensates, we see that the fourfold degeneracy at $\theta_{1,2}=\pi$ must be broken to a twofold degeneracy. 
The true ground states are either the pair $(++),(--)$ or $(+,-),(-,+)$. These two possibilities are physically distinct and lead to entirely different phase diagrams for the model (as well as different domain walls), as we will see. 

To find which is the right answer one has to integrate out the heavy fermion and compute the coefficient of the leading irrelevant operator that may lift the degeneracy among the four states~\eqref{fourgs}. Since the coefficient of such an irrelevant operator would have negative powers of the mass $M$, as long as $M$ is parametrically large, all it would do for the phase diagram is to deform it in a small neighborhood of $\te_{1,2}=\pi$. It is easy to see that the irrelevant operators $\int d^4x \left(\star \tr(F_1\wedge F_1)\right)^2$ and $\int d^4x \left(\star \tr(F_2\wedge F_2)\right)^2$ would not resolve the singularity at $\te_{1,2}=\pi$ because they do not allow for the two gauge theories to communicate. Hence, they do not break the accidental time reversal symmetry that acts separately on each of the $SU(N)$ gauge theory sectors. The leading operator that breaks this accidental time reversal symmetry is 
\begin{equation}\label{leadingop} \delta S = - {c\over M^4}\int d^4x \star \tr(F_1\wedge F_1) \star \tr(F_2\wedge F_2)~.\end{equation}
This operator is indeed invariant under the physical time reversal symmetry that acts on all the degrees of freedom simultaneously but it breaks the accidental time reversal symmetry that acts on the two sectors individually. 

The physics crucially depends on the sign of the coefficient $c$. The two options are depicted in \ref{2options}, where we draw the entire phase diagram for very large $M$. The anomalies and global inconsistency conditions we have previously considered are all consistent  with both scenarios. If $c>0$ then at $\te_{1,2}=\pi$ the true twofold degenerate vacua are $(++),(--)$ and if $c<0$ the true twofold degenerate ground states are $(+,-),(-,+)$.

One can compute $c$ from integrating out the fermion at one loop. This laborious computation is recorded in appendix \ref{1loop}. One finds that \begin{equation}\label{coeff}c=\frac{g_1^2g_2^2}{180(4\pi)^2}~.\end{equation}
We see that $c>0$ and hence the correct phase diagram for sufficiently large finite $M$ is given by the left diagram in \ref{2options}. One can  compute analytically the shape of the phase diagram to this order in $\Lambda/M$.

Let us now consider the domain walls in this limit of large $M$, starting from $M=\infty$. At $\theta_{1,2}=\pi$
we have four degenerate vacua, labeled as above by $(+,+),(+,-),(-,+),(-,-)$ and therefore we have several domain walls to consider. In fact there are only two distinct types of domain walls to consider: those where both signs flip and those where only one sign flips. 
Both are straightforward to analyze since the dynamics of the two gauge groups is decoupled. Depending on the orientation, if one sign flips we get $SU(N)_{\pm 1}$ TFT and if both signs flip we get the TFT $SU(N)_{\pm 1}\times SU(N)_{\pm1}$ (all the possible four combinations exist -- the signs are un-correlated between the two $SU(N)$ groups).
Actually the situation is more complicated because in case two signs flip the TFT arises from two separate $SU(N)$ gauge theory domain walls, and hence the distance between the walls, $d$, is unfixed. There is no force whatsoever between the two walls. The distance between the walls is described by a massless 2+1 dimensional scalar field, $d(x)$, uncharged under the gauge groups of the TFTs.

Now if we turn on a large finite mass $M$, as we argued before, we can integrate out the bi-fundamental field and since the coefficient~\eqref{coeff} is positive, at $\theta_{1,2}=\pi$ only the two vacua $(-,-),(+,+)$ survive. So let us consider the domain wall between these two vacua. For $M=\infty$ the theory on the wall consists of  the TFT $SU(N)_1\times SU(N)_1$, and, in addition, we have the center of mass mode and the additional massless degree of freedom describing the distance between the branes, $d(x)$. 

In the presence of the operator~\eqref{leadingop}  the two branes  attract each other. 
To see that, take the two branes to be widely separated. In the region between the branes we now have the state $(+,-)$ (or $(-,+)$).  But due to the operator~\eqref{leadingop} this has a higher energy density than the true ground state and hence it pays off for the branes to approach each other. This essentially means that there is a linear potential for the field $d(x)$ at large $d(x)$. This effect lifts the modulus $d(x)$, but, conjecturally, does nothing else.  It therefore seems reasonable to put forward a conjecture that the domain wall theory at large finite $M$ is  given by the TFT  $SU(N)_1\times SU(N)_1$ accompanied by the usual center-of-mass mode (and no other massless fields).

\section{The Massless Limit $M=0$}
\label{massless}
In the next section, we will study the phase diagram of the theory in the small mass limit. To prepare, in this section we study the massless theory $M=0$.
When the fermion mass vanishes, the classical Lagrangian is invariant under chiral rotations of the fermion 
\eql{chiraltransformation}{\psi\to e^{i\omega}\psi\ ,\ \tilde{\psi}\to e^{i\omega}\tilde{\psi}\ ,}
but this transformation is anomalous in the full theory. Indeed, a chiral transformation \eqref{chiraltransformation} shifts the two theta angles as
\eql{thetashift}{\te_{1,2}\to\te_{1,2}+2N\omega\ .}
We can learn from \eqref{thetashift} two interesting things:
\begin{enumerate}
\item The linear combination $\te_-\equiv\te_1-\te_2$ is physical since it is invariant under chiral transformations. The other linear combination $\te_+\equiv\onov{2}(\te_1+\te_2)$ is a redundant operator and the massless theory does not depend on it. In particular, the phase diagram can depend only on $\te_-$ and not on $\te_+$. 
\item There is a discrete non-anomalous subgroup of the chiral symmetry~\eqref{chiraltransformation} that shifts the $\te$ angles by integer multiple of $2\pi$. The subgroup corresponds to chiral transformations with $\omega=\frac{\pi k}{N}$ for some integer $k$. These transformations lead to a $\mathbb{Z}_{2N}$ symmetry of the full theory. More precisely, for even $N$, $\{-\psi,-\tilde{\psi}\}$ is gauge equivalent to $\{\psi,\tilde{\psi}\}$ so only a $\mathbb{Z}_N$ subgroup acts faithfully. For odd $N$, $\mathbb{Z}_{2N}$ acts faithfully. (For odd $N$, there are baryons which transform under the global symmetry with $k=N$.)
Below, the only transformations that will play a role in our analysis are the ones that act non-trivially on the fermion bilinear $\psi\tilde{\psi}$. 
Both for even and odd $N$, the fermion bilinear transforms under a $\mathbb{Z}_N$ chiral symmetry that acts on the fermion bilinear as
\eq{\psi\tilde{\psi}\to e^{2\pi ik/N}\psi\tilde{\psi}\ .}
\end{enumerate}

So far we have discussed some kinematics but now we have to discuss the long-distance physics of the theory. For large enough $N$, at $\theta_-=0$, we know that the  fermion bilinear expectation value $\vev{\psi\tilde{\psi}}$ is non-vanishing.  This follows from the planar equivalence of the bi-fundamental gauge theory with the $\mathcal{N}=1$ supersymmetric Yang-Mills theory. For some relevant discussions see~\cite{Strassler:2001fs,Gorsky:2002wt,Tong:2002vp,Kovtun:2004bz,Armoni:2005wta}.
Since the bilinear fermion operator transforms under  $\mathbb{Z}_N$, the condensate implies that there are actually $N$ vacua related by the symmetry. At large enough $N$ and $\theta_-=0$ these are all the vacua of the theory and they are trivially gapped.\footnote{In the literature there is extensive discussion~\cite{Strassler:2001fs,Gorsky:2002wt,Tong:2002vp,Kovtun:2004bz,Armoni:2005wta} of the following point: if the two gauge couplings are the same then the theory has an additional discrete $\mathbb{Z}_2$ symmetry interchanging the two gauge groups. Since ${\psi\tilde{\psi}}$ is invariant under this symmetry, one can then ask if this  $\mathbb{Z}_2$ symmetry is broken due to a nonzero order parameter such as $\vev{Tr F_1^2-Tr F_2^2}$.
This would have led to $2N$ vacua in the massless theory. We assume that such breaking does not occur. If it were to occur, there would need to be additional phase transitions at finite mass, further complicating the phase diagram. The present study reveals no evidence that the phase diagram needs to be more complicated than what we propose.}

\begin{figure}
	\begin{center}
	   	\includegraphics[width=0.8\textwidth, height=7cm]{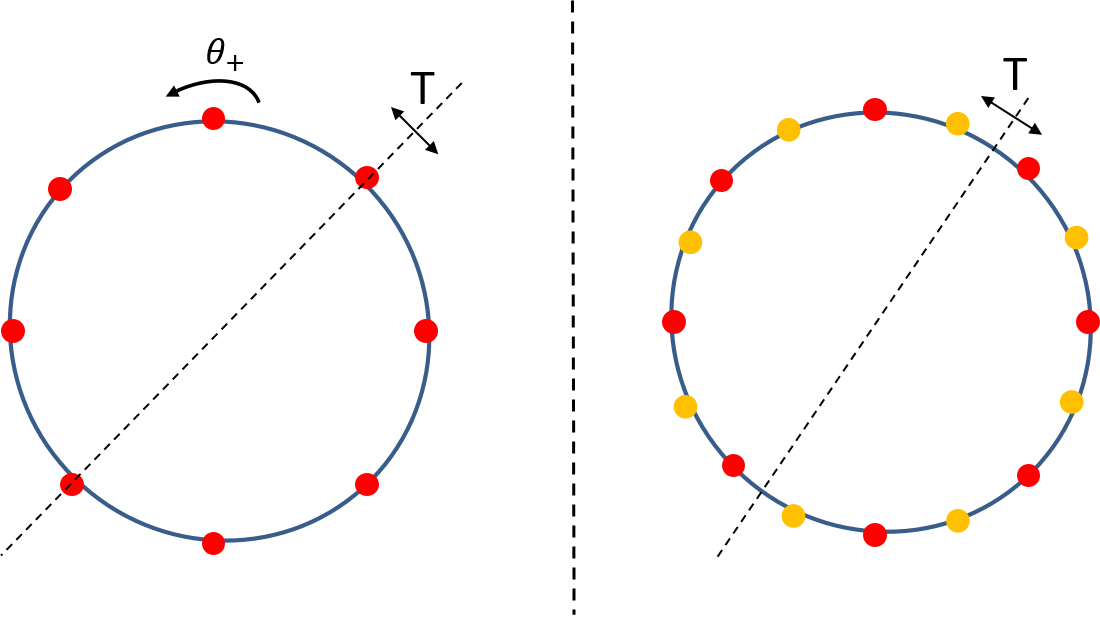}
  \caption{\small{On the left, the vacuum structure at $(\te_1,\te_2)=(0,0)$. The red dots are located at $\al_k=\frac{2\pi k}{N}$ with integer $k$. They represent the vacua with $\vev{\psi\tilde{\psi}}\sim e^{-i\al_k}$. For every one of these vacua, there is a combination of $\CT$ and chiral symmetry that acts as reflection around the corresponding dot and hence preserved by the vacuum. On the right, the vacuum structure at $(\te_1,\te_2)=(\pi/2,-\pi/2)$. Here, combinations of $\CT$ and chiral symmetry act as reflection between the dots and are broken by the vacuum. The yellow dots are new vacua located at $\al_k=\frac{2\pi k}{N}$ with half-integer $k$ and their existence is due to the spontanous breaking of $\CT$.  }}\label{redcircle}
	\vspace{10pt}
\end{center}

\end{figure}

For $\theta_1=\theta_2=0$, the fermion bilinear expectation value is given by 
\eql{conden}{\vev{\psi\tilde{\psi}}\sim e^{-i\al_k}~,\quad \al_k=\frac{2\pi k}{N}~,\ k=0,...,N-1} corresponding to our $N$ vacua. The spontaneously broken symmetry is generated by acting on the vacua as $\alpha_k\to\alpha_{k+1}$. 
These vacua can be visualized on a circle as the red dots in the left picture of figure \ref{redcircle}. 

Since the theory at $\theta_1=\theta_2=0$ admits time reversal symmetry, let us specify how it acts on these $N$ vacua. We can take $\CT$ to act on the fermion bilinear by complex conjugation, such that it takes $\al\to-\al$ which can be visualized as a reflection symmetry around $\al=0$. Therefore, we can combine $\CT$ with the broken chiral symmetry in such a way that each of the vacua is invariant under some anti-unitary symmetry and hence time reversal symmetry is unbroken.

What happens as we vary the $\te_-$ angle? (As we said, varying $\te_+$ is inconsequential at the massless limit; all it does is simply rotating the circle of figure \ref{redcircle} without any effect on the physics.) $\te_-$ is a physical parameter and is expected to influence the physics in a non-trivial way. 
Since the $\theta_-$ parameter is invariant under the $\mathbb{Z}_N$ symmetry, for small enough $\theta_-$ we will still have $N$ vacua. However, at $\theta_-=\pi$ we propose that there are $2N$ vacua due to the breaking of time reversal symmetry (which is a symmetry only at $\theta_-=0$ and $\theta_-=\pi$).
One piece of evidence for this doubling of the vacua comes from the mixed anomalies at $(\te_1,\te_2)=(\pi,0)$ that we already discussed. 

One can understand how this comes about in detail as follows: Denote by $\al_k(\te_1,\te_2)$ the phase of the bilinear condensate. The idea is to start at $\te_1=\te_2=0$ and follow the phases of the $N$ vacua. We can parameterize the phases with \eql{defphas}{\al_k(\te_1,\te_2)=\al_0(\te_1,\te_2)+\frac{2\pi k}{N}\ ,\ \al_0(0,0)=0\ ,\ k=0,...,N-1\ .}

We can continue following these vacua all the way to $\theta_-=\pi$, at which point there is again time reversal symmetry. Hence, we can try to understand how time reversal symmetry acts on these $N$ vacua. Consider for example the point $(\te_1,\te_2)=(\pi/2,-\pi/2)$. Time reversal symmetry acts by complex conjugation on the fermion bilinear and takes the theta angles to $(\te_1,\te_2)=(-\pi/2,\pi/2)$. To return to the original theory we can now perform the axial transformation~\eqref{chiraltransformation} with $\omega=\pi/2N$. Plugging this into~\eqref{thetashift} we see that this will transform us back to $(\te_1,\te_2)=(\pi/2,-\pi/2)$.  Therefore, time reversal acts at $(\te_1,\te_2)=(\pi/2,-\pi/2)$ as 
\eql{Treversepi}{\al\to-\al+\frac{\pi}{N}~.}
With similar considerations one can find how time reversal symmetry acts at $\te_-=\pi$ for generic $\theta_+$. The result is that for $(\te_1,\te_2)=(\pi/2+\te,-\pi/2+\te)$, time reversal acts as $\al\to-\al+\frac{\pi+2\te}{N}$.

Determining the phases of the bilinear condensate $\al_k(\te_1,\te_2)$ requires detailed knowledge about the dynamics at low energies. However, at least for $g_1=g_2$ things are under control. The reason is that when $g_1=g_2$ and $\te_1=-\te_2$, there is a new anti-unitary symmetry which is a combination of $\CT$ and the exchange symmetry $\mathcal{E}$ that exchanges the two gauge groups. This symmetry acts as $\al\to-\al$ and it is a symmetry that exists continuously as we go from $(\te_1,\te_2)=(0,0)$ to $(\te_1,\te_2)=(\pi/2,-\pi/2)$. This symmetry has no anomalies and there is no expectation that it is spontaneously broken. 

Since this symmetry acts by  $\al\to-\al$, it tells us that the vacuum at $\alpha=0$ cannot move as we increase $\theta_-$. Assuming this symmetry is unbroken, the vacua at $(\te_1,\te_2)=(\pi/2,-\pi/2)$ for $g_1=g_2$ are given by $\al_k=\frac{2\pi k}{N}$. 
Now, using \eqref{Treversepi}, we find that there must be $2N$ vacua given by 
\eql{condennew}{\vev{\psi\tilde{\psi}}\sim e^{-i\al_k}~,\quad \alpha_k=\pi k /N~,\ k=0,\ 1,\ ...,\ 2N-1~.}
Under the generator of the $\mathbb{Z}_N$ symmetry $\alpha_k\to \alpha_{k+2}$ and under time reversal symmetry $\alpha_k\to \alpha_{1-k}$. (The subscript of $\alpha$ is defined mod $2N$.) We depict the $2N$ vacua and their transformation properties in~\ref{redcircle}. 

Let us summarize and rephrase the main point: The theory with $g_1=g_2$ has an anti-unitary symmetry for all $\te_1=-\te_2$. In addition, there is a unitary (exchange) symmetry at $(\te_1,\te_2)=(0,0)$,
$(\te_1,\te_2)=(\pi/2,-\pi/2)$. We assume that the anti-unitary symmetry is never broken and the exchange symmetry is unbroken at  $(\te_1,\te_2)=(0,0)$. This leads to a determination of the fermion bilinear condensate. Then, due to the breaking of the exchange symmetry at $(\te_1,\te_2)=(\pi/2,-\pi/2)$, there is a doubling of the vacua and we have $2N$ ground states.

For $g_1$ and $g_2$ different but close enough, we expect the vacua at $(\pi/2,-\pi/2)$ to move a little bit from $\al_k=\pi k/N$, but we still expect to find $2N$ vacua. They are related by the action of \eqref{Treversepi}. 

At $\theta_-=\pi$ we expect $2N$ vacua for all $g_1$ and $g_2$. 
In general, it is not necessary for the $2N$ vacua to be visible in the fermion bilinear condensate. This is because there is another order parameter for the time reversal symmetry breaking at $\theta_-=\pi$, which is $\langle \star Tr F\wedge F\rangle$. Indeed, imagine that the fermion bilinear condensate at 
$\theta_-=\pi$ was $\al_k=\frac{\pi}{2N}+\frac{2\pi k}{N}$. These are mapped to themselves under \eqref{Treversepi} and hence the time reversal symmetry breaking would not be visible in the fermion bilinear condensate. 

In fact, in section \ref{chirallag} we analyze the phase diagram when $M=0,\ \Lambda_1\gg\Lambda_2$ and find that when $\te_-=\pi$ there are indeed $2N$ vacua as generally predicted, but the fermion bilinear condensate only takes $N$ values. This can be explained precisely as outlined above:  changing $g_1,g_2$ shifts the vacua away from their original values at $g_1=g_2$ (which we determined above), until at some critical $g_1(g_2)$, the vacua reach the fixed point of \eqref{Treversepi}, $\al_k=\frac{\pi}{2N}+\frac{2\pi k}{N}$ and then the fact that there are $2N$ vacua can be understood by also considering the order parameter $\langle \star Tr F\wedge F\rangle$.

Let us now summarize our proposal for the dynamics of the massless theory for generic $\Lambda_1,\Lambda_2$.  At all $\theta_-\neq\pi$ there are $N$ vacua related by the spontaneously broken $\mathbb{Z}_N$ symmetry. However, at $\theta_-=\pi$ new $N$ vacua appear (this is a first order transition) and they are related to the old $N$ vacua by time reversal symmetry. Whether or not these new $N$ vacua are visible in the fermion bilinear condensate $\vev{\psi\tilde{\psi}}$ is a more complicated question. We have argued that this is the case for $\Lambda_1,\Lambda_2$ that are sufficiently close to each other, but in section  \ref{chirallag} we will see that for hierarchical strong coupling scales this is not the case and the splitting of the $N$ vacua to $2N$ vacua can be only understood by including additional order parameters.

The scenario we propose here appears to be the simplest one in which the small mass and the large mass limits can be connected smoothly without any additional phase transitions, as will be shown in the next section. In addition, the claims here are consistent with planar equivalence, anomalies (and global inconsistency conditions), the 't Hooft limit (which we analyze later), and with the more detailed analysis of section  \ref{chirallag}.

\begin{figure}
	\vspace{10pt}
	\includegraphics[width=1\textwidth, height=5.5cm]{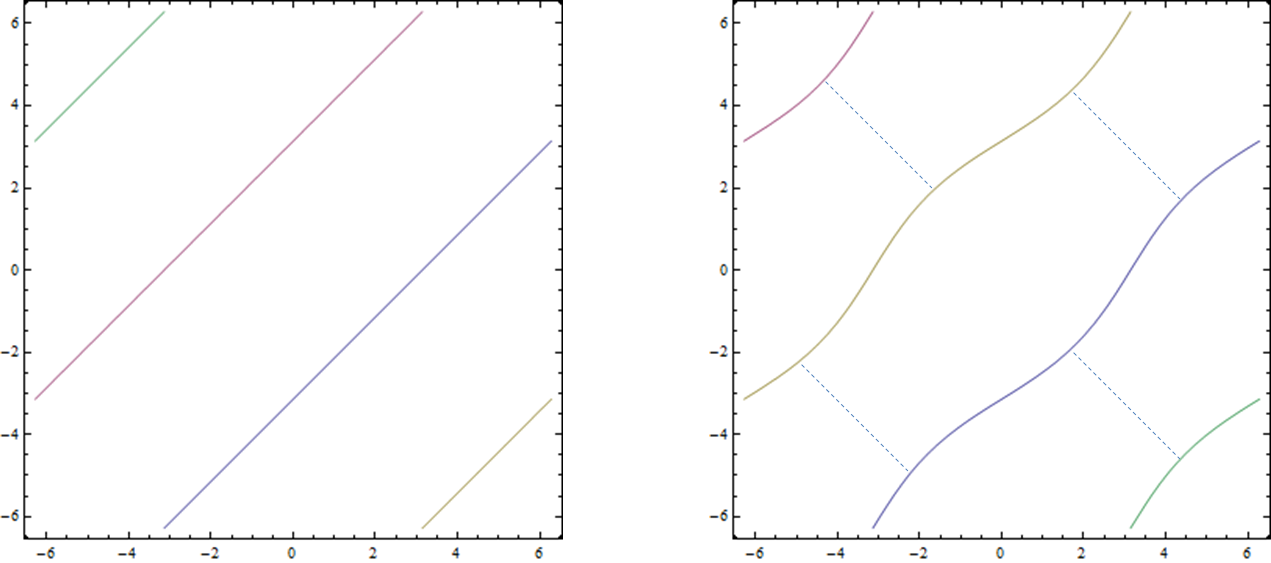}
	\caption{\small{The left figure represents the phase diagram in the $M=0$ case. Nothing depends on $\theta_+$ since it is not a parameter in the massless theory. The right figure represents the phase diagram for some small finite mass.  At $M=0$, the discrete chiral symmetry leads to an $N$-fold degeneracy in each region away from the transitions (at which there are $2N$ vacua), while at $M>0$, the vacuum is unique away from the transition lines. The dashed lines disappear as we take $M=0$. }}\label{smallM}
	\vspace{10pt}
\end{figure}

\section{Small $M\neq 0$}
\label{smallm}
In this section we will add a small mass term $\delta\lag=M \psi\tilde{\psi}+c.c.$ as a perturbation to the massless theory studied in the previous section. The mass term breaks the discrete chiral symmetry explicitly and therefore also the $\mathbb{Z}_N$ degeneracy.
As before, shifts in $\te_+$ are related to shifts in $\al$, but now $\te_+$ is a physical parameter because the mass term is not invariant under chiral transformations. The physical effect of $\theta_+$ comes from the dependence of the potential on $\te_+$, $V\sim -M\cos\left(\al\right)$.
 Let us first consider the case of $g_1=g_2$. As argued in the previous section, when $g_1=g_2$ and the theory is massless symmetry arguments determine the phases of the fermion condensate in the vacua, $\al_k(\te_+)=\frac{\pi k+\te_+}{N}$. The states with even $k$ are the true vacua for $-\pi+4\pi \mathbb{Z}<\te_-<\pi+4\pi \mathbb{Z}$, and the states with odd $k$ are the true vacua for $\pi+4\pi \mathbb{Z}<\te_-<3\pi+4\pi \mathbb{Z}$. The lines of $\te_-=\pi +2\pi \mathbb{Z}$ are the transition lines in which the $2N$ states are degenerate.

In the vicinity of $\te_-=0$, only the $N$ states with even $k$ are important. Then, the change in the energy of the $k$-th vacuum ($k=0,...,N-1$) is given by \eql{pert}{\delta E_k\sim-M\cos\left(\frac{\pi k+\te_+}{N}\right)\ .}
For generic values of $\te_+$ the vacuum is unique. However, we clearly see a transition line whenever $\te_+=\pi \mod 2\pi \mathbb{Z}$.
Similarly, in the vicinity of $\te_-=2\pi$, only the states with odd $k$ are important, and we will have transition lines whenever $\te_+=0 \mod 2\pi \mathbb{Z}$.

The analysis of the vicinity of $\theta_-=\pi$ is a little more complicated since there are more states to consider. We can work to linear order in $M$ and in $\delta\theta_-=\theta_--\pi$ but exactly in $\theta_+$. The potential on the space of nearly-degenerate $2N$ vacua, parameterized by the integer $k=0,...,2N-1$ as in~\eqref{condennew}, is given by 
\eql{linanal}{V(k) = -M\cos\left(\frac{\pi k+\te_+}{N}\right)+(-1)^ka \delta \theta_-\ .}
The second term in \eqref{linanal} reflects the fact that for $\delta\te_->0$, the odd $k$ states have less energy than the even $k$ states, and for $\delta\te_-<0$, the even $k$ states have less energy. Here $a$ is some unknown positive constant. This formula captures correctly, to linear order in $M$ and $\delta \theta_-$, the energy differences between the vacua.
This potential reproduces the features we have studied in detail for $M=0$ in the previous section.
At $\delta\theta_-=0$ there is clearly a two fold degeneracy at $\theta_+=\pi/2\mod \pi \mathbb{Z}$.

In general, there is a smooth curve $\theta_+(\delta\theta_-)$ which describes a line of phase transitions. 
We have just seen that $\theta_+(0)=\pi/2$. This line must at some point intersect the other line of phase transitions, which we have described above, where $\theta_+=\pi$. At that intersection there must be three degenerate vacua. We can find this intersection point from~\eqref{linanal}.
Indeed, let us plug $\theta_+=\pi$ and evaluate the energy of the three vacua corresponding to $k=-2,-1,0$: 
\eql{threecands}{E_{k=-1}=-M-a\delta\te_-  ~,\quad E_{k=-2}=E_{k=0}= -M\cos\left(\frac{\pi}{N}\right)+a \delta \theta_-~.}
These three states are degenerate for $\delta \theta^*_-=-\frac{M}{2a}(1-\cos\left(\frac{\pi}{N}\right))$. Note that $\delta\theta^*_-<0$ at this point.
By inspecting~\eqref{linanal} it is not hard to convince oneself that this is the only point with three vacua.  
We therefore see that there is a curve of first order phase transitions around $\theta_-=\pi$, curving a little bit upwards for small $M$, and at $\delta\theta_-^*$ intersecting the line $\theta_+=\pi$. The phase diagram for small $M$ is depicted  in figure \ref{smallM}.

Now we can move on to the more general case where $g_1\neq g_2$. In this case, we do not have complete control over the phases of the bilinear condensate. However, as explained in the previous section, for $g_1$ and $g_2$ sufficiently close by, the $2N$ vacua of the massless theory at $\theta_-=\pi$ are visible in the fermion condensate and so we can proceed to our analysis of the massive theory even without complete knowledge of the fermion bilinear condensate. 

Let us start from $(\te_1,\te_2)=(0,0)$, in which there are $N$ vacua at (see equation~\ref{defphas}) $\al_k(0,0)=\frac{\pi k}{N}$ with $k=0,\ 2,\ 4,\ ...$. 
As we now crank up $\theta_-$ and allow for some generic value of $\theta_+$ in the massless theory, these $N$ vacua are located at
\eql{startpo}{\al_k(\te_+,\te_-)=\al_-(\te_-)+\frac{\pi k+\te_+}{N}\ ,\ \al_-(0)=0\ .}
New $N$ vacua must appear in the massless theory 
at $\theta_-=\pi$ and for $g_1$ and $g_2$ sufficiently close to each other, they are all visible in the bilinear fermion condensate. For this to happen, as explained in the previous section, we need to assume that $\al_-(\pi)<\frac{\pi}{2N}$. Indeed, if this is the case,  time reversal symmetry at $\te_-=\pi,\ M=0$ acting on the fermion bilinear leads to $N$ new values of the condensate. 
The $2N$ vacua can be parametrized by
\eq{\al_k=(-1)^k\al_-(\te_-)+\frac{\pi k+\te_+}{N}\ ,\ k=0,1,2,...\ .}
For $M=0$, the even $k$ states are the true vacua for $-\pi+4\pi \mathbb{Z}<\te_-<\pi+4\pi \mathbb{Z}$, and the odd $k$ states are the true vacua for $\pi+4\pi \mathbb{Z}<\te_-<3\pi+4\pi\mathbb{Z}$. The lines of $\te_-=\pi +2\pi \mathbb{Z}$ are the transition lines in which the $2N$ states are degenerate.

As in the previous case, turning on a small mass is equivalent to adding the potential $V\sim-M\cos\left(\al\right)$. The energy contribution to each of the states is
\eql{diffcouplingmass}{\delta E_k\sim -M\cos\left(\frac{\pi k+\te_++(-1)^kN\al_-(\te_-)}{N}\right)\ .}
In the vicinity of $\te_-=0$, the even $k$ states are the important ones. \eqref{diffcouplingmass} leads to a transition line whenever $\te_++N\al_-(\te_-)=\pi\ {\rm mod}\ 2\pi\mathbb{Z}$.
In the same way, in the vicinity of $\te_-=2\pi$, the odd $k$ states are the important ones.  \eqref{diffcouplingmass} leads to a transition line whenever $\te_+-N\al_-(\te_-)=0\ {\rm mod}\ 2\pi\mathbb{Z}$.

Close to $\te_-=\pi$, we can write in a similar fashion to \eqref{linanal}
\eq{V(k)= -M\cos\left(\frac{\pi k+\te_++(-1)^kN\al_-(\te_-)}{N}\right)+(-1)^ka \delta \theta_-\ .}

We can look for the triple point in which the three states with $k=0,-1,-2$ are degenerate. The degeneracy between $k=0$ and $k=-2$ happens when $\te_++N\al_-=\pi$. $k=-1$ is also degenerate with them when
\eq{M\cos\left(\frac{\pi}{N}\right)-M\cos\left(2\al_-(\te_-)\right)=2a\delta\te_-\ .}
To leading order in both $M$ and $\delta\te_-$, we can replace $\al_-(\te_-)\simeq \al_-(\pi)$, and get
\eq{\delta\te_-=\frac{M}{2a}\left[\cos(\pi/N)-\cos(2\al_-(\pi))\right]\ .}

This leads to an interesting conclusion. We can see that as we vary $\al_-(\pi)$ from $0$ to $\frac{\pi}{2N}$, the triple point moves continuously from $\left(\frac{3\pi}{2}-\epsilon, \frac{\pi}{2}+\epsilon\right)$ to $\left(\pi,0\right)$ with $\epsilon=\frac{M}{4a}\left[1-\cos(\pi/N)\right]$. Accordingly, the phase diagram transform from diagram (b) of figure \ref{cases} to diagram (c) of figure \ref{cases}. 

This is an important and crucial consistency check, establishing the continuous relation between the behavior of the theory with similar dynamical scales and hierarchical dynamical scales. As we will see in the next section, the behaviour of the theory with widely different dynamical scales can be studied using altogether different techniques; the fact that we obtain diagram (c) of figure \ref{cases} is therefore really encouraging. Moreover, in the next section we show that if we take $\Lambda_1/\Lambda_2\to \infty$ then $\al_-(\te_-)=\frac{\te_-}{2N}$. This precisely leads to  diagram (c) of figure \ref{cases}.

%

%
%

\section{Hierarchical Dynamical Scales and the Chiral Lagrangian} 
\label{chirallag}
In this section we will study the limit in which one of the strong scales is much larger than the mass and the other strong scale $M\ll\Lambda_1$, and $\Lambda_2\ll\Lambda_1$. The way to study this limit is first take $M=\Lambda_2=0$ and write the chiral Lagrangian that arises from confinement of the first gauge group at scale $\Lambda_1$. To this chiral Lagrangian, we can add a mass term as a small perturbation and gauge the $SU(N)$ global symmetry with gauge coupling $g_2$. We will first introduce $g_2$ on top of the dynamics of the first gauge group and then add a mass. 

\newpage 
\begin{center}{\underline {Massless Case}}\end{center}

 The dynamics of the massless theory has been presented  already in section \ref{massless}. Here we will re-derive these results for $M=0$ from the chiral Lagrangian point of view. The analysis that goes into the chiral Lagrangian is entirely different from the considerations in~\ref{massless} and hence further reinforce the whole picture.

As explained above, for energies $\Lambda_2\ll E\ll \Lambda_1$ the gauge coupling of the second group is still very weak. The strongly coupled gauge group leads to chiral symmetry breaking, with the low energy theory described by special unitary matrices $U\in SU(N)$ and the action being as usual $$\mathcal {L}\sim \partial U \partial U^{-1}+\cdots ~.$$
This action by itself is invariant under the $SU(N)_L\times SU(N)_R$ global symmetry, $U\to VUV'$ with $V\in SU(N)_L$ and $V'\in SU(N)_R$.

Now we need to weakly gauge the diagonal subgroup
$SU(N)\subset SU(N)_L\times SU(N)_R$. Therefore the action becomes 
\begin{equation}\label{dgauge}\mathcal {L}\sim D U D U^{-1}+{1\over g_2^2}Tr(F_2^2)+i{\theta_2^{EFF}\over 8\pi^2 } Tr(F_2\wedge F_2)\cdots ~,\end{equation}
 with $D$ now the appropriate covariant derivative such that the action is invariant under the local gauge transformations $U\to g(x) U g(x)^{-1}$ with $g(x)\in SU(N)$. (We will consider the meaning of the theta term $\theta_2^{EFF}$ below. Note that the subscript $2$ in $\theta_2$ and $F_2$ stands for the fact that this is the field strength associated to the second gauge group. We hope this notation would not cause any confusion.)

First, let us ask what are the remaining global symmetries. 
We started with global symmetry $SU(N)_L\times SU(N)_R$ and gauged the diagonal subgroup. 
The normalizer of the diagonal subgroup is given by the set of elements $(A, B)\in SU(N)_L\times SU(N)_R$ such that for all $C$, there exists some $D$ such that  $$ (ACA^{-1},BC^{-1}B^{-1}) = (D,D^{-1})~.$$
For this to be true it must be that $A^{-1}B$ is a scalar $SU(N)$ matrix. Hence, the normalizer is given by $N$ copies of the group $SU(N)$. The remaining global symmetry is the quotient of the normalizer by the gauge symmetry and hence it is simply $\mathbb{Z}_N$. 

Indeed, the global symmetry of the gauged Lagrangian~\ref{dgauge} is manifestly generated by multiplying the matrix $U$ by the scalar $e^{2\pi i /N}$.
We therefore arrived at the anticipated result that the global symmetry of the model~\eqref{dgauge} is $\mathbb{Z}_N$, acting by multiplying $U$ with a scalar $SU(N)$ matrix, in agreement with our expectations from the microscopic point of view. 

Actually, the full massless bi-fundamental model has a  bigger symmetry. There is a $U(1)$ baryon symmetry and (for odd $N$) the chiral symmetry is $\mathbb{Z}_{2N}$ and not $\mathbb{Z}_N$. It is however a common phenomenon that some symmetries are lost in the effective theory -- this is because the symmetries that are lost only act on heavy objects. Indeed, the $\mathbb{Z}_2\subset \mathbb{Z}_{2N}$ that is invisible for odd $N$ as well as baryon number symmetry (in fact the $\mathbb{Z}_2$ is contained in the baryon symmetry) are realized on Skyrmions.

The discrete global symmetry  $\mathbb{Z}_N$ acting by multiplying $U$ with a scalar $SU(N)$ matrix cannot be broken by perturbative or non-perturbative effects in the coupling $g_2$.

What are the classical vacua? Since classically there is no potential for $U$, the classical vacua are given by the space of all possible constant matrices $U$ modulo gauge transformations. The space of special unitary matrices $U$ modulo the gauge transformations $gUg^{-1}$ is of course \begin{equation}\label{vacspace}S\left[U(1)^N\right]/S_N~.\end{equation}
Namely, it is the space of diagonal matrices with eigenvalues on the circle, modulo re-arrangements of the eigenvalues. The constraint that the determinant is one is implemented by the $S$ operation in the numerator of~\eqref{vacspace}.

The classical theory therefore has an $N-1$ dimensional space of vacua. It is important to realize that the space~\eqref{vacspace} has singularities when the action of $S_N$ has fixed points. Indeed, that is the case when some eigenvalues coincide. In the event that some eigenvalues coincide, there is an unbroken non-Abelian gauge symmetry. At a generic point of~\eqref{vacspace} the massless gauge fields are Abelian.

The next task is to determine the fate of these vacua~\eqref{vacspace} in the quantum theory. There are perturbative as well as non-perturbative effects in $g_2$. First, there is a potential that is induced on the space \eqref{vacspace} at one loop. This potential lifts the moduli space except for precisely those points, where the gauge symmetry is completely unbroken. Namely the following vacua remain after including the one-loop effective potential in $g_2$: 
\eql{afterone}{U=e^{2\pi i k \over N} 1_{N\times N}~.}
These $N$ vacua are related by the discrete $\mathbb{Z}_N$ global symmetry and hence at one loop the  $\mathbb{Z}_N$ global symmetry is spontaneously broken. 

The statement that the one loop effective potential lifts the moduli space~\eqref{vacspace} and only leaves the discrete vacua~\eqref{afterone} is familiar from a rather different context: A famous concern is that technicolor theories may cause vacuum misalignment, namely, that the chiral condensate would not preserve the electromagnetic gauge symmetry. In the 80s it was proven that this does not occur and the vacuum is aligned~\cite{Witten:1983ut}. This is why the charged pions have a larger mass than the neutral pions.
For us that means that the chiral condensate, $U$, would not spontaneously break the diagonal gauge symmetry. This is consistent only if the matrix $U$ is a scalar matrix and hence we have our $N$ vacua~\eqref{afterone}. We conclude that at one loop the moduli space is lifted and we get spontaneous breaking of the global $\mathbb{Z}_N$ symmetry.

Expanding around each of these vacua, and taking into account the one-loop correction, one finds that the physics is that of a massless $SU(N)$ gauge theory (with gauge coupling $g_2$ at that scale) coupled to an adjoint scalar with a positive mass squared. In order for the adjoint scalar to be weakly coupled at the scale of its mass, we further assume \begin{equation}\label{adjointmass} m^2_{Adj}~\sim g_2^2\Lambda_1^2~\gg \Lambda_2^2~.\end{equation}

With this assumption, since the adjoint scalar has a mass much above the strong coupling scale $\Lambda_2$, we can safely integrate it out and we thus remain with pure $SU(N)$ Yang-Mills theory in each of the $N$ vacua.
If $\theta_2^{EFF}=\pi \mod 2\pi$ then each of these vacua is further split into two vacua by the condensation of $\star Tr( F_2\wedge F_2)$.
What remains is to relate $\theta_2^{EFF}$ to the microscopic $\theta_1,\theta_2$. This is done by simply doing a chiral transformation in the original model such that the theta angle of the first gauge group vanishes. Hence we have
$$\theta_2^{EFF}=\theta_2-\theta_1=-\theta_-~.$$

In summary, we see that as expected, the dynamics is independent of $\theta_1+\theta_2$ but it depends on $\theta_-\equiv \theta_1-\theta_2$. For generic $\theta_-$ we get $N$ trivially gapped vacua related by the discrete $\mathbb{Z}_N$ axial symmetry. For $\theta_-=\pi$ we instead find $2N$ vacua, where the doubling of the number of vacua is due to the additional spontaneous breaking of time reversal symmetry. 

Let us now compare these results with the ones obtained in section \ref{massless}.
The two methods lead to topologically the same phase diagram, which is given by $N$ states at generic $\te_-$, and $2N$ states at $\te_-=\pi+2\pi\mathbb{Z}$ in which time reversal is spontaneously broken. We can make the relation even sharper by comparing the possible vacuum expectation values of $\tr\psi\tilde{\psi}$ with the possible vacuum expectation values of $Tr U$. 
As argued above, $Tr U$ can take $N$ values given by $\vev{Tr U}_k=Ne^{2\pi i k/N}$ with $k=0,...,N-1$. This is true even when $\te_-=\pi$ where the vacua are doubled due to time reversal breaking by the order parameter $\vev{Tr(F_2\wedge F_2)}$. Time reversal breaking is not visible in the order parameter $TrU$. 
Notice that in this analysis we took the limit $\Lambda_1\gg g_2\Lambda_1\gg\Lambda_2$ which enabled us to integrate out the massive pions while the second gauge group is still weakly coupled. This picture is consistent with section \ref{massless}, where we argued that for similar strong coupling scales the time reversal breaking is visible in the fermion bilinear order parameter while for hierarchically different strong coupling scales the time reversal breaking is not visible in the fermion bilinear order parameter.\footnote{Previously we defined the object $\alpha_-(\theta_-)$ as in equation~\ref{startpo}. We have computed the condensate $U$ in the particular choice that $\theta_+=-\theta_-/2$ (since we rotated away the theta angle of the first gauge group to the second gauge group) and comparing with~\ref{startpo} we can therefore infer that $\alpha_-(\theta_-) = \theta_-/2N$.  This is precisely as was anticipated at the end of section~\ref{smallm}. }

\subsection{Massive Quarks}

Let us now consider the regime of $\Lambda_2,M \ll \Lambda_1$  where $M$ is the mass of the bi-fundamental Dirac fermion and it is taken to be positive and real without loss of generality. (We will see that we will need a slightly more refined ordering of the scales to be able to analyze the system.)
Solving for the dynamics of the first gauge group we arrive at essentially the same model as above, only deformed by a mass term in the chiral Lagrangian: 
\begin{equation}\label{dgaugeM}\mathcal {L}\sim Tr( D U D U^{-1})-\left(Me^{-i\theta_1/N} TrU+c.c.\right)+{1\over g_2^2}Tr(F_2^2)+i{\theta_2-\theta_1\over 8\pi^2} Tr (F_2\wedge F_2)\cdots ~,\end{equation}
The mass term completely breaks the $\mathbb{Z}_N$ discrete axial symmetry as expected. We study the model in the limit $\Lambda_2,M \ll \Lambda_1$ but in addition, we want to think about $M$ as a relatively small perturbation on top of what we have found at $M=0$ in the previous subsection. Indeed, as explained in the previous subsection, there is a contribution to the mass of the adjoint scalar from loops proportional to $g_2^2$~\eqref{adjointmass}, and we want those to dominate over $M$. So we will assume, in addition, that $M\ll g_2\Lambda_1$. With these assumptions, we can treat $M$ as a small perturbation over the $N$ (or $2N$ at $\theta_-=\pi$) gapped vacua we previously had.

Now the classical space of vacua is given by the space of $SU(N)$ matrices modulo gauge transformations but in addition we have to reside at the minima of the potential, namely we have to minimize $\left(Me^{i\theta_1/N} TrU+c.c.\right)$. First, we diagonalize the matrix $U$ using the gauge symmetry and write $$U=diag(e^{i\psi_1},...,e^{i\psi_N})~,\quad \sum_i \psi_i = 0 \ mod \ 2\pi~.$$
As in the previous section, the induced potential on the space of $\{\psi_i\}$ at one loop in the  gauge coupling $g_2$ lifts the entire space except for the points for which $\psi_i=\frac{2\pi k}{N}$ for some integer $k$. 
Then the potential on the space of the $\psi_i$ that comes from the mass term is given by 
$$-M\sum_i \cos\left(\psi_i-{\theta_1\over N}\right)=-MN\cos\left(\frac{2\pi k-\te_1}{N}\right)~.$$
The result is that for $|\theta_1|<\pi$ the ground state is given by $\psi_i=0$ while for $\theta_1=\pi$ the ground state is two-fold degenerate, due to the degeneracy between the solutions
$\psi_i=0$ and $\psi_i={2\pi\over N}$. Shifting $\theta_1$ by $2\pi$ can be offset by shifting the $\psi_i$ by $2\pi/N$ and hence this pattern repeats itself.

So far we have described the vacua of the matrix $U$. The $U$ fluctuations around these vacua are all massive with mass of order $g_2\Lambda_1$ (neglecting the small corrections proportional to $M$). 
We see that the heavy degrees of freedom in the matrix $U$ have two degenerate vacua for $\theta_1=\pi \mod 2\pi \mathbb{Z}$ and one vacuum otherwise. 
Since there are still light fields described by the gauge field with field strength $F_2$, our analysis is not complete: the vacua may be further split and/or some degeneracies may be removed. 

To approach this problem we have to understand the symmetries of the model~\eqref{dgaugeM} in more detail. Most importantly, we have to understand how time reversal symmetry acts. 
Let us start by completely ignoring the gauge field $A_2$. Then we have the Lagrangian 
\begin{equation}\label{dgaugeMi}\mathcal {L}\sim Tr( \partial U \partial U^{\dagger})-\left(Me^{-i\theta_1/N} TrU+c.c.\right)+\cdots ~.\end{equation}
In addition, it will be important to remember that in the $\cdots$, there is a Wess-Zumino term. 

To understand how time reversal symmetry may act, consider the following two possibilities: 
$$\CT_1: U(t,x)\to U^\dagger(-t,x)~,\quad \CT_1: U^\dagger (t,x)\to U(-t,x)~,$$
or
$$\CT_2: U(t,x)\to U(-t,x)~,\quad \CT_2: U^\dagger(t,x)\to U^\dagger (-t,x)~. $$
The kinetic term is invariant under both $\CT_1,\CT_2$ transformations. 
The mass term transforms as follows (we think about $\theta_1$ as a spurion -- if it is invariant then it means that the symmetry is not broken explicitly): 
$$\CT_1: Me^{i\theta_1/N}\to Me^{i\theta_1/N} ~,$$
$$\CT_2: Me^{i\theta_1/N}\to Me^{-i\theta_1/N}~.$$
Let us now investigate how the Wess-Zumino term transforms under $\CT_1$ and $\CT_2$. First, consider the Hermitian matrix $i U^\dagger \partial U $. If the derivative acts in the space direction we see that 
$\CT_1: i U^\dagger \partial U \to -i U \partial U^\dagger=i\partial U  U^\dagger $ and $\CT_2:  i U^\dagger \partial U \to -i U^\dagger \partial U$. As a result, if we consider the trace
$$i^n\epsilon_{\mu_1,\mu_2,...,\mu_n}Tr( U^\dagger \partial^{\mu_1} U U^\dagger \partial^{\mu_2} U\cdots  U^\dagger\partial^{\mu_n} U)$$ 
then, for $n=5$, under $\CT_1$ it is odd while under $\CT_2$ it is even.\footnote{One has to remember that at least one of the derivatives acts in the time direction, which is why there is an additional minus sign.}
If we denote the coefficient of the Wess-Zumino term by $C_{WZ}$, we therefore see that as a spurion 
$$\CT_1: C_{WZ}\to -C_{WZ}~,$$
$$\CT_2: C_{WZ}\to C_{WZ}~.$$
Finally, another useful symmetry is the $\mathbb{Z}_N$ symmetry, which is broken by the mass term but unbroken by the Wess-Zumino term. It therefore acts by 
$$\mathbb{Z}_N: C_{WZ}\to C_{WZ}~,\quad Me^{i\theta_1/N}\to Me^{i\theta_1/N}e^{2\pi i / N}~.$$

One corollary of the above is that for some special values of $\theta_1$ the theory has unbroken time reversal symmetry. To see that we may combine $\CT_2$ with the $\mathbb{Z}_N$ symmetry. This leads to a theory invariant under time reversal only if $\theta_1$ and $-\theta_1$ are the same mod $2\pi$. In other words, only if $\theta_1=\pi \ {\rm mod} \ 2\pi$ or if $\theta_1=0 \ {\rm mod} \ 2\pi$. This is of course precisely what we expect from the microscopic model. 

We can now couple the theory back to the dynamical gauge field $A_2$. Path integrating over $U$ we find some effective action for $A_2$. Indeed, consider the system at energies \eql{scales}{\Lambda_2 \ll E\ll g_2\Lambda_1~.}
The effective field theory is described by the gauge field $A_2$. 

That effective action is constrained by gauge invariance and by the symmetries above. 
We are most interested in the term in that effective action $Tr(F_2\wedge F_2)$. It is time reversal odd under both $\CT_1$ and $\CT_2$. 
Since $A_2$ is invariant under the $\mathbb{Z}_N$ symmetry, so is $Tr(F_2\wedge F_2)$. Let us denote the coefficient of that term which arises from integrating out the heavy fields by 
$\Delta\theta$. The symmetry properties above imply that it must take the form 
$$\Delta\theta\sim C_{WZ}f(\theta_1)~,$$
with $f(\theta_1)$ an odd, $2\pi$ periodic function of the argument (this function does not have to be differentiable because the heavy fields described by the degrees of freedom in $U$ sometimes have more than one ground state). This is  consistent with the $\mathbb{Z}_N$ action as well as the time reversal actions $\CT_1$,$\CT_2$. 
The effective theta angle of the theory at scales~\eqref{scales} is given by
\eq{\theta_{ir}=\theta_2-
\theta_1+
\Delta \theta~.}

In the regime $M\ll g_2\Lambda_1$, the effects of the tree-level mass for $U$, $\left(Me^{-i\theta_1/N} TrU+c.c.\right)$, are small and hence we can expand $\Delta \theta$ in powers of $M\sin(\theta_1/N)$. From our spurion analysis we know that there must be some insertions of $\theta_1$ in order to satisfy the selection rules.  As a result, we can estimate that  \eql{esteff}{\Delta\theta\sim M/\Lambda_2\ll 1~.}
(It may be parametrically even smaller than that. And there are possibly additional factors of $g_2$ and $N$.) 
If we shift $\te$ by $2\pi$ we jump to a new vacuum for the heavy field $U$. That means that at $\theta_1=\pi \mod 2\pi$ the function $\Delta\theta$ need not be smooth. For this reason, for instance, a linear dependence on $M\sin(\theta_1/N)$ is not in contradiction with the $2\pi$ periodicity of the physics under $\theta_1\to \theta_1+2\pi$. 

What we learn is that $\Delta\theta$ is parametrically small. Therefore, neglecting $\Delta\theta$, the phase diagram can be described as follows (see figure (c) of \ref{cases}).
For $\theta_1\neq \pi \ {\rm mod}\ 2\pi$, the heavy fields reside in a single vacuum, and the $A_2$ gauge field leads to a splitting of that vacuum if $\theta_2-\theta_1=\pi \ {\rm mod} \ 2\pi$. For $\theta_1=\pi \ {\rm mod}\ 2\pi$, the heavy fields have two vacua. Neglecting $\Delta\theta$, each of these is split into two vacua by the light fields if $\theta_2-\theta_1=\pi \ {\rm mod} \ 2\pi$. Therefore, for $\theta_1=\pi \ {\rm mod}\ 2\pi$ and $\theta_2=0 \ {\rm mod}\ 2\pi$ we have four vacua. 

As we have seen in our discussion in subsection 3.2, this kind of four-fold degeneracy should not be exact and it should generically be resolved into two trivalent vertices. Indeed, this is precisely what the correction $\Delta\theta$ achieves.

\begin{figure}
	\vspace{10pt}
\begin{center}
	\includegraphics[width=0.5\textwidth, height=6cm]{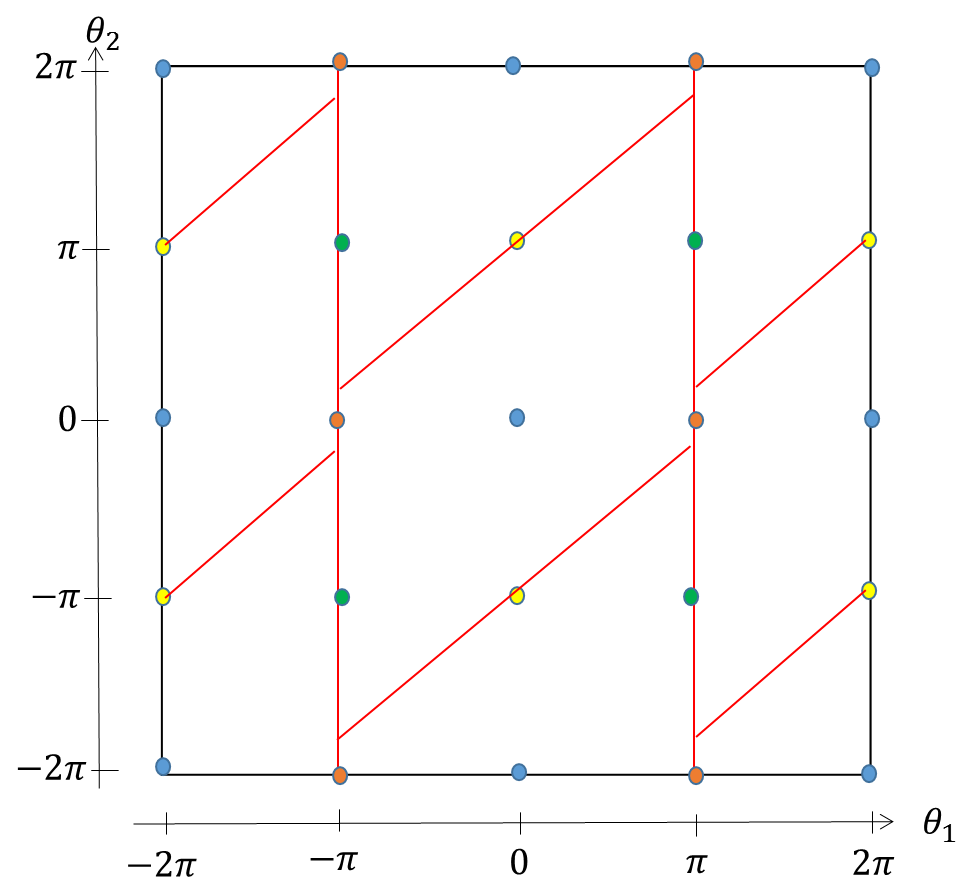}
	\caption{\small{Schematic plot of the phase diagram in the limit $\Lambda_1\gg M,\Lambda_2$.}}\label{chiraldiagram}
\end{center}
	\vspace{10pt}
\end{figure}

There are two distinct ways for the diagram to be resolved upon including $\Delta\theta$. Which is the correct one depends on the sign of $\Delta\theta$, which we did not compute. We merely point out though, that for the appropriate sign the phase diagram will take the form of figure~\ref{chiraldiagram}, which is topologically the same as what we have found in the various limits we have studied so far. Even without computing $\Delta \theta$ explicitly, we know that the point $(\te_1,\te_2)=(0,\pi)$ (and points related to it by periodicity) has a twofold degenerate vacuum because at this point all the corrections are even under $\CT$ and hence it must be that $\Delta\theta=0$. However, for $\theta_1=\pi$, there is spontaneous breaking of $\CT$ by the heavy fields and hence in the two vacua the sign of $\Delta\theta$ is opposite. This leads to a resolution of the singularity at $\theta_1=\pi,\theta_2=0$ (and points related to it by the periodicity) according to the sign of $\Delta\theta$.

It would be very nice to complete this analysis and verify that the topology of the phase diagram is indeed the same as in the other limits we have studied. However, even without that, the analysis we have carried out here leads to several nontrivial consistency checks and corroborates our proposal.

\section{$SU(N_1)\times SU(N_2)$}
\label{diffrank}
In this section we will explain how the results of this paper change if we take the two gauge groups to have different ranks. 
We will only consider asymptotically free theories such that $N_2<\frac{11N_1}{2}$ (without loss of generality, we take here $N_2\geq N_1$). 

We can investigate the phases of the model in the various limits we have considered before for $N_1=N_2$.
The results for $M\gg\Lambda_{1,2}$ are analogous to those for the $N_1=N_2$ case. This will result in a phase diagram that looks qualitatively the same as the left figure in \ref{2options}.
The small mass limit, on the other hand, looks very different since the physical $\te$ angle at the massless limit depends on the ranks $N_{1,2}$. 
 Let us define
\eq{N=gcd(N_1,N_2)\ ,\ n_1=\frac{N_1}{N}\ ,\ n_2=\frac{N_2}{N}\ .}
 When $M=0$, the chiral transformations $\psi\to e^{i\omega}\psi\ ,\ \tilde{\psi}\to e^{i\omega}\tilde{\psi}$ shift the theta angles as
\eq{\te_1\to\te_1+2N_2\omega\ ,\ \te_2\to\te_2+2N_1\omega\ .}
We therefore have a $\mathbb{Z}_N$ non-anomalous chiral symmetry with $\omega=\frac{\pi k}{N}$.\footnote{As in section \ref{massless}, we consider only the $\mathbb{Z}_N$ subgroup as it acts non-trivially on the fermion bilinear $\psi\tilde{\psi}$.} The physical $\te$ angle is the combination which is invariant under chiral transformations \eql{tephys}{\te_{phys}=n_1\te_1-n_2\te_2\ .}

Unlike the equal rank case, in general the phase diagram is topologically different in the small and large mass limits. Matching the two limits leads to non-trivial behavior at intermediate values of the mass $M$. 

To understand the nature of the topology changes in the phase diagram, in this section we will be discussing the theory only at the large $N_{1,2}$ limit. (On the way we will also reproduce some of the claims in the previous sections for the equal rank case.) Furthermore, we will assume that $N=gcd(N_1,N_2)\gg 1$.

As we will see, the theory undergoes a duality cascade as we reduce $M$, similar in structure to the Klebanov-Strassler duality cascade~\cite{Klebanov:2000hb}, see also~\cite{Strassler:2005qs} for a review. (Note that the cascade that we encounter here concerns with the behavior of the theory as we change the mass, not the renormalization group scale.)

\subsection{Large $N$ and a Duality Cascade}
\label{seccascade}
In this section we will construct the phase diagrams in the large $N$ limit~\cite{tHooft:1973alw}.
A simplification that occurs in the large $N$ limit is that for a fixed mass and gauge couplings, the energy of the vacuum state can be expanded in $\te_{1,2}/N$ (see~\cite{Witten:1980sp} and also \cite{Witten:1998uka} and references therein). Keeping only the leading terms, the vacuum energy is a quadratic function in the $\te$ angles
\eq{E_{(0,0)}=E_0+a_1\te_1^2+a_2\te_2^2+b\te_1\te_2+O(1/N^2)\ .}
The four parameters $E_0, a_{1,2}, b$ depend on the mass and the dynamical scales. $E_0$ scales like $N^2$ while $a_{1,2}$ and $b$ are $O(1)$ functions of the mass and strong coupling scales.
In order to satisfy the $2\pi$ periodicities of the $\te$ angles, there must exist vacua with energies
\eql{energies}{E_{(k_1,k_2)}=E_0+a_1(\te_1+2\pi k_1)^2+a_2(\te_2+2\pi k_2)^2+b(\te_1+2\pi k_1)(\te_2+2\pi k_2)\ .}
These are the famous branches of meta-stable states that are commonplace in large $N$ Yang-Mills theory.
For instance, there is a degeneracy between $E_{(0,0)}$ and $E_{(k_1,k_2)}$ when
\eq{(2k_1a_1+k_2b)\te_1+(2k_2a_2+k_1b)\te_2=-2\pi(k_1^2a_1+k_2^2a_2+k_1k_2b)\ .}

In the large mass limit, we know that $a_1\sim\Lambda_1^4\ ,\ a_2\sim\Lambda_2^4\ ,\ b\to 0^-$. It is straightforward to solve for the curves where vacua become degenerate and construct the phase diagram of the large mass limit.  
The result is universal and qualitatively independent of the ranks $N_1,N_2$. It coincides with the left figure in \ref{2options} and figure~\ref{cases}(a).\footnote{For instance, focusing on the fundamental domain $(\theta_1,\theta_2)\in [0,2\pi]
\times [0,2\pi] $, the curve that separates the phase $k_1=0,k_2=0$ from $k_1=0,k_2=-1$ is given by
$b \theta_1+2a_2(\theta_2-\pi)=0$ and one can similarly find all the other curves separating the ground states in the fundamental domain. Then, connecting the curves one finds the phase diagram in the large mass limit, and using that $b$ is small and negative we find the claimed result.}

\begin{center}  $$\underline{ N_1=N_2 }$$\end{center}
We will start from the equal rank case in order to warm up with the large $N$ techniques and to reproduce some of the results of the previous sections. When $M=0$, the energy depends only on the combination $\te_{phys}=\te_1-\te_2$ and takes the form $E_{(0,0)}=E_0+a(\te_1-\te_2)^2$ such that $a_i=a\ ,\ b=-2a$. 
First we see that $E_{(0,0)}=E_{(k,k)}$ for every $k$. Similarly, $E_{(1,0)}=E_{(1+k,k)}$ for every $k$. This degeneracy reflects the fact that there are $N$ vacua at generic values of the $\theta$ angles. 

The transition between the two families happens when $\te_2-\te_1=\pi$ in agreement with the results of the previous sections (see the left figure~\ref{smallM}). Indeed, at $\te_2-\te_1=\pi$ we expect $2N$ vacua.

 Now we can turn on a small mass which breaks the degeneracy between $E_{(k,k)}$ for different $k$. Parametrizing $a_i=a+\epsilon_i\ ,\ b=-2a+\epsilon$, the transition line between the two lowest lying states in this family in the fundamental domain is given by
\eq{E_{(0,0)}=E_{(-1,-1)}\Rightarrow (2\epsilon_1+\epsilon)\te_1+(2\epsilon_2+\epsilon)\te_2=2\pi(\epsilon_1+\epsilon_2+\epsilon)\ .}
This line always goes through the point $(\pi,\pi)$. Its slope in the $(\te_1,\te_2)$ plane is $S=-\frac{2\epsilon_1+\epsilon}{2\epsilon_2+\epsilon}$. 
$S$ can take various values as a function of these parameters $\epsilon_i$, $\epsilon$. For instance, if the two gauge groups have the same strong coupling scale then $\epsilon_1=\epsilon_2$ and then the slope is $S=-1$, which is exactly consistent with our general considerations, see for instance figure~\ref{cases}(b). 

\newpage

\begin{center}  $$\underline{ N_1\neq N_2 }$$\end{center}

\begin{figure}
	\vspace{10pt}
         	\includegraphics[width=0.8\textwidth, height=4.5cm]{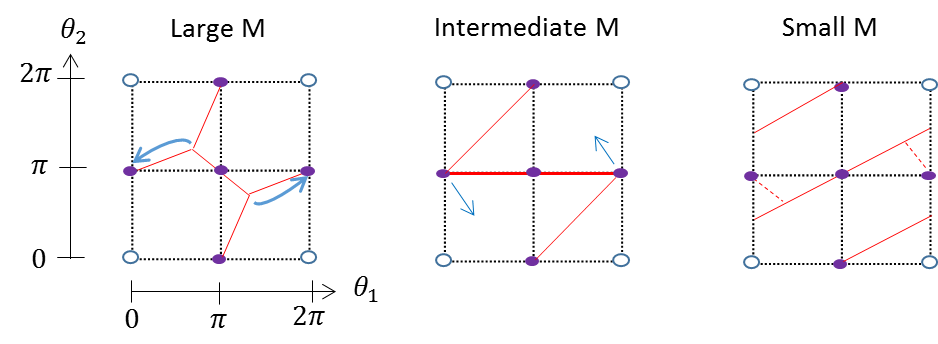}
  \caption{\small{An example of the expected flow from large $M$ to small $M$ for $n_1=1$, $n_2=2$. The empty circles represent the points at which $\cT$ is preserved in the vacuum. The purple circles represent the points at which $\cT$ is spontaneously broken. The blue arrows indicate the direction in which the intersection points move as we decrease $M$. As we see, there must be some intermediate $M$ for which there is a point with fourfold degeneracy. The dashed lines disappear when $M=0$. }}\label{flow}
	\vspace{10pt}
\end{figure}

To understand some of the new features that appear for different ranks, let us study the theory with $N_1=N$, $N_2=2N$ and $N\to\infty$. The main difference is that now the physical theta angle at $M=0$ is $\te_{phys}=\te_1-2\te_2$. Hence, the energies at $M=0$ take the form
\eq{E_{(0,0)}=E_0+a(\te_1-2\te_2)^2\ ,\ E_{(k_1,k_2)}(\te_1,\te_2)=E_{(0,0)}(\te_1+2\pi k_1, \te_2+2\pi k_2)\ .}
The phase diagram of the massless theory already presents us with an interesting puzzle.
One can minimze the energy as usual and one finds that in the fundamental domain there are four regions, corresponding to the rightmost figure~\ref{flow} (ignoring the dashed lines). These four regions, from the top left corner to the bottom right corner correspond to $(k_1=0,k_2=-1)$, $(k_1=k_2=-1)$, $(k_1=k_2=0)$, $(k_1=-1,k_2=0)$, and in all cases we can shift $k_1\to k_1+2p, k_2\to k_2+p$ for any integer $p$.
A striking fact about this phase diagram is that the regions $(k_1=0,k_2=-1)$ and $(k_1=k_2=0)$ are not separated by a phase transition. This is in contrast with the large mass region, where they are always separated by a phase transition. 
This already means that there must be some interesting change in topology at some finite mass. When we add a small mass perturbation the phase diagram that we find corresponds to the rightmost figure~\ref{flow} but now including the dashed lines. This phase diagram has six regions, which further shows that there must be some change in topology at some finite mass, since the large mass phase diagram has only four regions (see the leftmost figure~\ref{flow}). 

We can follow the transition line between $E_{(0,0)}$ and $E_{(-1,-1)}$ from large to small mass. 
This transition line is given by
\eq{(2a_1+b)\te_1+(2a_2+b)\te_2=2\pi(a_1+a_2+b)\ .}
For any value of $a_{1,2}$ and $b$, this line goes through the point $(\pi,\pi)$. The slope in the $(\te_1,\te_2)$ plane is $S=-\frac{2a_1+b}{2a_2+b}$.
For large mass, $b\to 0$ and the slope is negative (since the topological susceptibility at $\theta=0$ in the pure YM theory is positive), as in the leftmost figure~\ref{flow}. For zero mass (or very small small), the slope is $S=\onov{2}$. In order to connect the two limits, there should be some $M^*(\Lambda_1,\Lambda_2)$ such that the slope is zero (which happens at $2a_1=-b$). This point is distinguished because there are four ground states at $(0,\pi)$. These different limits and the associated phase diagrams are depicted in figure \ref{flow}. At $M^*(\Lambda_1,\Lambda_2)$ a topology change of the phase diagram occurs.

A curious fact about the point $M^*$ is that upon the redefinitions 
\eql{redef}{(\te_1',\te_2')=(\te_1-\te_2,\te_2)\ ,\ (k_1',k_2')=(k_1-k_2,k_2)\ ,}
we find the energy spectrum
\eql{dualspectrum}{E_{(k_1',k_2')}=E_0+a_1(\te_1'+2\pi k_1')^2+ (a_2-a_1)(\te_2'+2\pi k_2')^2\ .}
This is exactly the energy spectrum of two decoupled YM theories $SU(N_1')\times SU(N_2')$! (By spectrum, we mean the energy spectrum of the infinitely many meta-stable states.)
This is not entirely unexpected in light of the discussion above, since at $M^*$ we have a quartic vertex in the phase diagram, so the change of variables~\eqref{redef} simply signifies that the diagram can be rotated to look like that of two decoupled YM theories at $M^*$.

What happens when we continue to lower the mass below $M^*$? the slope should become positive when we take $M<M^*$, which means that $2a_1+b=-\epsilon<0$. Plugging it into \eqref{energies} with the redefinitions \eqref{redef} we find the spectrum
\eq{E_{(k_1',k_2')}=E_0+a_1(\te_1'+2\pi k_1')^2+ (a_2-a_1-\epsilon)(\te_2'+2\pi k_2')^2-\epsilon(\te_1'+2\pi k_1')(\te_2'+2\pi k_2')\ .}
It looks like the spectrum of $SU(N_1')\times SU(N_2')$ coupled to a new (dual) bi-fundamental fermion with very large mass. 
What are the ranks of the two groups? This can be found by looking at the physical theta angle at the massless limit. In the original theory, we had $\te_{phys}=\te_1-2\te_2$. Using \eqref{redef} we find $\te_{phys}'=\te_1'-\te_2'$ which is the physical theta angle for $N_1'=N_2'$. Furthermore, agreement between the global symmetries of the two theories is achieved if $N_1'=N_2'=N$. 
This resemblance between the two theories suggests a duality between $SU(N)\times SU(2N)$ and theta angles $(\te_1,\te_2)$ to $SU(N)\times SU(N)$ with theta angles $(\te_1-\te_2,\te_2)$: 
\begin{equation}\label{dualitybi}SU(N)\times SU(2N)+{\rm bi-fundamental} \longleftrightarrow SU(N)\times SU(N)+{\rm bi-fundamental}\end{equation}
In the 't Hooft limit all these transitions are first order and one may ask about the nature of the duality~\ref{dualitybi}. First of all, it extends beyond the ground state: it also applies to the meta-stable states that are characteristic to large $N$ gauge theories. Second, it conjecturally applies to the domain walls that appear at the respective first order phase transitions. Since these domain walls often carry nontrivial degrees of freedom at long distances, the duality has nontrivial infrared content. 

Let us now generalize this cascade-like duality to general ranks.
We start from an $SU(N_1)\times SU(N_2)$ gauge theory with a bi-fundamental fermion in the 't Hooft limit and $N=gcd(N_1,N_2)\gg 1$ (where we define $n_i=N_i/N$ and assuming without loss of generality $n_1<n_2$).
 At large mass there are four regions and the  slope separating $k_1=k_2=-1$ from $k_1=k_2=0$ is negative exactly as in the previous cases. At zero mass, the physics can depend only on $\te_{phys}=n_1\te_1-n_2\te_2$. This generally leads to a completely different phase diagram when compared to the large mass region. There must exist special values $M^*$ where the topology of the phase diagram changes. 
 In general, several such points where the topology of the phase diagram changes are necessary. 
 
We have found a general scenario that allows for the topology to transmute as necessary. 
As we lower the mass, a first critical $M=M^*(\Lambda_1,\Lambda_2)$ is encountered. At $M^*$  the slope of the line separating $k_1=k_2=-1$ from $k_1=k_2=0$  is zero and we have $2a_1=-b$.\footnote{In the same way, if $n_1>n_2$, there is a critical $M^*$ for which $2a_2=-b$ and the slope is infinite.} At this point, we can use the redefinition \eqref{redef} and write the spectrum as in \eqref{dualspectrum} which is again the spectrum of decoupled $SU(N_1')\times SU(N_2')$. In order to find the new ranks, we compare the physical theta angle at the massless limit. In the original theory it was $\te_{phys}=n_1\te_1-n_2\te_2$. In terms of the new theta angle it is $\te'_{phys}=n_1\te_1'-(n_2-n_1)\te_2'$. Comparing also the global symmetries, we find that the new ranks are $N_1'=N_1\ ,\ N_2'=N_2-N_1$.
From this point, we can continue lowering the mass below $M^*$ and work with the dual theory which is $SU(N_1)\times SU(N_2-N_1)$ with massive bi-fundamental fermion. 
We can continue with the same procedure until we reach the $SU(N)\times SU(N)$ theory which is the last step in the cascade. This is illustrated in figure \ref{cascade}. At every step of the cascade we simply take the gauge group with the bigger rank (if it exists) and subtract the smaller rank from it.

Some comments about this duality cascade are in order:
\begin{enumerate}
\item One may worry that $a_2-a_1<0$ such that the spectrum in \eqref{dualspectrum} is not physical. However, there is a restriction on the parameters $a_{1,2},\ b$ due to Witten \cite{Witten:1998uka}, who argued that the vacuum energy of Yang-Mills should be minimal at $\te=0$. Applying the same reasoning to our theory, we find the conditions\footnote{The inequalities are expected to be saturated only for $M=0$, where one combination of the $\te$- angles becomes unphysical.} \eql{mintest}{a_{1,2}\geq 0\ ,\ 4a_1a_2\geq b^2\ .} Using conditions \eqref{mintest} for $b=-2a_1$ we find that $a_2-a_1\geq 0$.

\item Using \eqref{redef}, the energies of the $SU(N_1)\times SU(N_2)$ theory can be written as
\eq{E_{(k_1',k_2')}=E_0+a'_1(\te'_1+2\pi k_1')^{2}+a'_2(\te'_2+2\pi k_2')^{2}+b'(\te_1'+2\pi k_1')(\te_2'+2\pi k_2')\ ,}
with \eq{a'_1=a_1\ ,\ a'_2=a_1+a_2+b\ ,\ b'=2a_1+b\ .}
The spectrum \eqref{dualspectrum} with arbitrary values of $a_1',a_2',b'$ doesn't always come from a physical microscopic $SU(N_1)\times SU(N_2-N_1)$ theory.   In other words, the three equations \eql{3to3}{a_1'=a_1'(M',\Lambda_1',\Lambda'_2)\ ,\ a_2'=a_2'(M',\Lambda_1',\Lambda'_2)\ ,\ b'=b'(M',\Lambda_1',\Lambda'_2)\ ,} don't necessarily have a solution with real and non-negative $M',\ \Lambda_1',\ \Lambda_2'$. Indeed, the duality maps the parameters of the two theories such that $M=M^*$ is mapped to $M'=\infty$ and $M=0$ is mapped to $M'=0$.
\item Ordinarily, infrared dualities apply for gapless theories or in the vicinity of a gapless point. Here, the statement of the duality~\ref{dualitybi} is somewhat weaker. We claim that the phases of the theories match, the phase transition lines match, and in addition, all the branches of meta-stable vacua match. Conjecturally, also all the domain walls in the various first order phase transitions match (more on this below). Since these domain walls may carry low energy degrees of freedom, the statement is highly nontrivial. 

\item Domain walls: Since there is a mapping between the two phase diagrams, it is very natural to conjecture that also the domain walls of the two dual theories can be mapped to each other. In the 't Hooft limit it makes sense to discuss the domain walls separating the true (degenerate) vacua as well as domain walls separating the true vacua from the meta-stable ones. It would be nice to understand this in detail!

\item Finally, it is natural to ask what happens when $N_1,N_2$ are not necessarily large and, even more conservatively,  what happens when $gcd(N_1,N_2)$ is not necessarily large. The topological arguments hold also for finite $N_{1,2}$, so we can expect that at least qualitatively, the phase diagrams behave in a similar way. We leave these questions to the future. In particular, the theory with $gcd(N_1,N_2)=1$ seems analogous to $N_f=1$ QCD and it would be nice to understand its dynamics better.

\end{enumerate}

\section*{Acknowledgments}

We would like to thank  A. Armoni, I. Klebanov, O. Mamroud, N. Seiberg, M. Strassler, and C.~Vafa for comments. A.K is supported by the ERC STG grant 335182. Z.K. is supported in part by the Simons Foundation grant 488657 (Simons Collaboration on the Non-Perturbative Bootstrap). Any opinions, findings, and conclusions or recommendations expressed in this material are those of the authors and do not necessarily reflect the views of the funding agencies.

\appendix
\section{1-loop contribution to $\tr(F\wedge F)\tr(F\wedge F)$ }
\label{1loop}
\begin{figure}
	\vspace{10pt}
         	\includegraphics[width=1\textwidth, height=7cm]{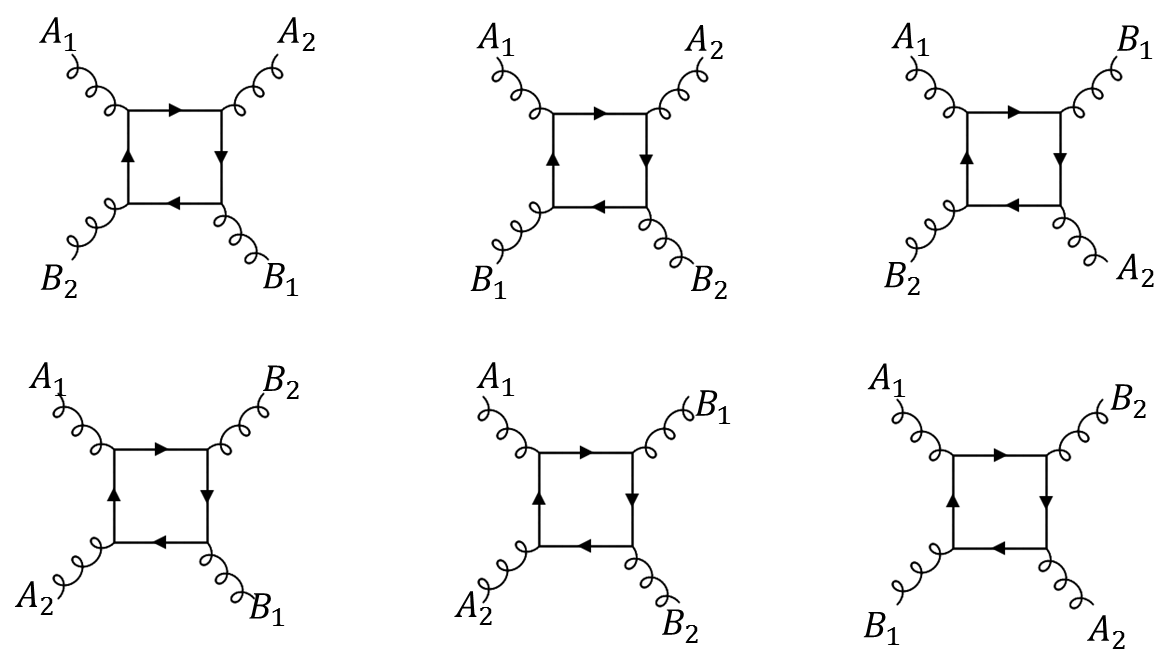}
  \caption{\small{The six diagrams that contributes to the 4 gluon interaction. Here $A_i\equiv A_{\mu_i}^{a_i}(p_i)$ and $B_i\equiv B_{\nu_i}^{b_i}(q_i)$ with $q_2=-p_1-p_2-q_1$.   }}\label{Feynman}
	\vspace{10pt}
\end{figure}

In this appendix we will compute explicitly the 1-loop contribution to the $(F_1\wedge F_1)(F_2\wedge F_2)$ term.
Lets denote the first gauge field by $A$ and the second by $B$ for the ease of notations. The fermionic action is\footnote{In this appendix we use the Dirac notations for the fermion.}
\eq{i\bar{\psi}_{\al\be}\gamma^\mu\partial_\mu\psi_{\be\al}+\bar{\psi}_{\al\be}\gamma^\mu A_\mu^aT^a_{\be\gamma}\psi_{\gamma\al}-\bar{\psi}_{\al\be}\gamma^\mu \psi_{\be\gamma}B_\mu^aT^a_{\gamma\al}-M\bar{\psi}_{\al\be}\psi_{\be\al}\ .}
We are looking for a correction to the action of the form
\eq{\delta S=C\int d^4x\tr(T^{a_1}T^{a_2})\tr(T^{b_1}T^{b_2})\epsilon^{\mu_1\mu_2\mu_3\mu_4}\epsilon^{\nu_1\nu_2\nu_3\nu_4}\partial_{\mu_3}A_{\mu_1}^{a_1}\partial_{\mu_4}A_{\mu_2}^{a_2}\partial_{\nu_3}B_{\nu_1}^{b_1}\partial_{\nu_4}B_{\nu_2}^{b_2}\ ,}
where $C=C(M,g_i)$. We didn't write the full gauge invariant quantity, just the term which is quartic in the gluons for simplicity. The contribution to the energy is $\delta S$. If $C<0$, it is favorable to increase $(F_A\wedge F_A)(F_B\wedge F_B)$ and therefore in the vacuum they have the same sign. If $C>0$, in the vacuum $F_A\wedge F_A$ and $F_B\wedge F_B$ have opposite signs. In Fourier space,
\eq{\delta S=&C\int d^4x\int \frac{d^4p_1d^4p_2d^4q_1d^4q_2}{(2\pi)^{16}}\tr(T^{a_1}T^{a_2})\tr(T^{b_1}T^{b_2})\epsilon^{\mu_1\mu_2\mu_3\mu_4}\epsilon^{\nu_1\nu_2\nu_3\nu_4}\\&p_{1,\mu_3}A_{\mu_1}^{a_1}(p_1)p_{2,\mu_4}A_{\mu_2}^{a_2}(p_2)q_{1,\nu_3}B_{\nu_1}^{b_1}(q_1)q_{2,\nu_4}B_{\nu_2}^{b_2}(q_2)e^{-i(p_1+p_2+q_1+q_2)x}\\
=&-C\int \frac{d^4p_1d^4p_2d^4q_1}{(2\pi)^{12}}\tr(T^{a_1}T^{a_2})\tr(T^{b_1}T^{b_2})\epsilon^{\mu_1\mu_2\mu_3\mu_4}\epsilon^{\nu_1\nu_2\nu_3\nu_4}\\&p_{1,\mu_3}A_{\mu_1}^{a_1}(p_1)p_{2,\mu_4}A_{\mu_2}^{a_2}(p_2)q_{1,\nu_3}B_{\nu_1}^{b_1}(q_1)(p_1+p_2+q_1)_{\nu_4}B_{\nu_2}^{b_2}(-p_1-p_2-q_1)\ .}
The tree level result from this interaction term will be
\eq{P_{\mu_1\mu_2\nu_1\nu_2}^{a_1a_2b_1b_2}=-iC\tr(T^{a_1}T^{a_2})\tr(T^{b_1}T^{b_2})\epsilon^{\mu_1\mu_2\mu_3\mu_4}\epsilon^{\nu_1\nu_2\nu_3\nu_4}p_{1,\mu_3}p_{2,\mu_4}q_{1,\nu_3}(p_1+p_2+q_1)_{\nu_4}\ ,}
where
\eq{P_{\mu_1\mu_2\nu_1\nu_2}^{a_1a_2b_1b_2}=\vev{A_{\mu_1}^{a_1}(p_1)A^{a_2}_{\mu_2}(p_2)B^{b_1}_{\nu_1}(q_1)B^{b_2}_{\nu_2}(-p_1-p_2-q_1)}\ .}
Therefore
\eq{&-\epsilon_{\mu_1\mu_2\mu_3\mu_4}\epsilon_{\nu_1\nu_2\nu_3\nu_4}\frac{\partial^4P_{\mu_1\mu_2\nu_1\nu_2}^{a_1a_2b_1b_2}}{\partial p_1^{\mu_3}\partial p_2^{\mu_4}\partial q_1^{\nu_3}\partial p_1^{\nu_4}}\\
&=iC\tr(T^{a_1}T^{a_2})\tr(T^{b_1}T^{b_2})\epsilon_{\mu_1\mu_2\mu_3\mu_4}\epsilon_{\nu_1\nu_2\nu_3\nu_4}\left[\epsilon^{\mu_1\mu_2\mu_3\mu_4}\epsilon^{\nu_1\nu_2\nu_3\nu_4}+\epsilon^{\mu_1\mu_2\nu_4\mu_4}\epsilon^{\nu_1\nu_2\nu_3\mu_3}\right]\\
&=720iC\tr(T^{a_1}T^{a_2})\tr(T^{b_1}T^{b_2})\ .}
There are six diagrams at 1-loop order that contribute to this correlation function. They appear in figure \ref{Feynman}. The three diagrams on the second line are the same as the three diagrams on the upper line just with opposite orientation. Because there is an even number of fermionic propagators, they just give a factor of two. Therefore, it is enough to study the three diagrams on the upper line.

Instead of computing the full diagram, we will extract the relevant part by computing 
\eq{-\epsilon_{\mu_1\mu_2\mu_3\mu_4}\epsilon_{\nu_1\nu_2\nu_3\nu_4}\frac{\partial^4P_{\mu_1\mu_2\nu_1\nu_2}^{a_1a_2b_1b_2}}{\partial p_1^{\mu_3}\partial p_2^{\mu_4}\partial q_1^{\nu_3}\partial p_1^{\nu_4}}_{p_1=p_2=q_1=0}\ .}
Start with the first diagram on \ref{Feynman}. 
the contribution is
	\eq{&\vev{A^{a_1}_{\mu_1}(p_1)A^{a_2}_{\mu_2}(p_2)B^{b_1}_{\nu_1}(q_1)B^{b_2}_{\nu_2}(q_2)}=-g_1^2g_2^2\tr(T^{a_1}T^{a_2})\tr(T^{b_1}T^{b_2})\int\frac{d^4k}{(2\pi)^4}\\
		&\tr\left[\gamma_{\mu_1}\frac{\slashed{k}+M}{k^2-M^2}\gamma_{\mu_2}\frac{\slashed{k}+\slashed{p}_2+M}{(k+p_2)^2-M^2}\gamma_{\nu_1}\frac{\slashed{k}+\slashed{p}_2+\slashed{q}_1+M}{(k+p_2+q_1)^2-M^2}\gamma_{\nu_2}\frac{\slashed{k}-\slashed{p}_1+M}{(k-p_1)^2-M^2}\right]}
	
We will start by computing  \eq{\epsilon_{\mu_1\mu_2\mu_3\mu_4}\epsilon_{\nu_1\nu_2\nu_3\nu_4}\frac{\partial^4}{\partial p_1^{\mu_3}\partial p_2^{\mu_4}\partial q_1^{\nu_3}\partial p_1^{\nu_4}}P(p_1=p_2=q_1=0)\ .}
First we will take derivatives of the integrand
\eq{&\frac{\partial}{\partial p_2^{\mu_4}}\tr\left[\gamma_{\mu_1}\frac{\slashed{k}+M}{k^2-M^2}\gamma_{\mu_2}\frac{\slashed{k}+\slashed{p}_2+M}{(k+p_2)^2-M^2}\gamma_{\nu_1}\frac{\slashed{k}+\slashed{p}_2+\slashed{q}_1+M}{(k+p_2+q_1)^2-M^2}\gamma_{\nu_2}\frac{\slashed{k}-\slashed{p}_1+M}{(k-p_1)^2-M^2}\right]\\
	&=\tr\left[\gamma_{\mu_1}\frac{\slashed{k}+M}{k^2-M^2}\gamma_{\mu_2}\frac{\gamma_{\mu_4}}{k^2-M^2}\gamma_{\nu_1}\frac{\slashed{k}+\slashed{q}_1+M}{(k+q_1)^2-M^2}\gamma_{\nu_2}\frac{\slashed{k}-\slashed{p}_1+M}{(k-p_1)^2-M^2}\right]\\
	&-2(k+q_1)_{\mu_4}\tr\left[\gamma_{\mu_1}\frac{\slashed{k}+M}{k^2-M^2}\gamma_{\mu_2}\frac{\slashed{k}+M}{k^2-M^2}\gamma_{\nu_1}\frac{\slashed{k}+\slashed{q}_1+M}{((k+q_1)^2-M^2)^2}\gamma_{\nu_2}\frac{\slashed{k}-\slashed{p}_1+M}{(k-p_1)^2-M^2}\right]\\
	&+\tr\left[\gamma_{\mu_1}\frac{\slashed{k}+M}{k^2-M^2}\gamma_{\mu_2}\frac{\slashed{k}+M}{k^2-M^2}\gamma_{\nu_1}\frac{\gamma_{\mu_4}}{(k+q_1)^2-M^2}\gamma_{\nu_2}\frac{\slashed{k}-\slashed{p}_1+M}{(k-p_1)^2-M^2}\right]\\
	&-2k_{\mu_4}\tr\left[\gamma_{\mu_1}\frac{\slashed{k}+M}{k^2-M^2}\gamma_{\mu_2}\frac{\slashed{k}+M}{(k^2-M^2)^2}\gamma_{\nu_1}\frac{\slashed{k}+\slashed{q}_1+M}{(k+q_1)^2-M^2}\gamma_{\nu_2}\frac{\slashed{k}-\slashed{p}_1+M}{(k-p_1)^2-M^2}\right]}
Now,
\eq{&\frac{\partial}{\partial q_1^{\nu_3}}[...]=\\
	&\tr\left[\gamma_{\mu_1}\frac{\slashed{k}+M}{k^2-M^2}\gamma_{\mu_2}\frac{\gamma_{\mu_4}}{k^2-M^2}\gamma_{\nu_1}\frac{\gamma_{\nu_3}}{k^2-M^2}\gamma_{\nu_2}\frac{\slashed{k}-\slashed{p}_1+M}{(k-p_1)^2-M^2}\right]\\
	&-2k_{\nu_3}\tr\left[\gamma_{\mu_1}\frac{\slashed{k}+M}{k^2-M^2}\gamma_{\mu_2}\frac{\gamma_{\mu_4}}{k^2-M^2}\gamma_{\nu_1}\frac{\slashed{k}+M}{(k^2-M^2)^2}\gamma_{\nu_2}\frac{\slashed{k}-\slashed{p}_1+M}{(k-p_1)^2-M^2}\right]\\
	&-2\eta_{\nu_3\mu_4}\tr\left[\gamma_{\mu_1}\frac{\slashed{k}+M}{k^2-M^2}\gamma_{\mu_2}\frac{\slashed{k}+M}{k^2-M^2}\gamma_{\nu_1}\frac{\slashed{k}+M}{(k^2-M^2)^2}\gamma_{\nu_2}\frac{\slashed{k}-\slashed{p}_1+M}{(k-p_1)^2-M^2}\right]\\
	&-2k_{\mu_4}\tr\left[\gamma_{\mu_1}\frac{\slashed{k}+M}{k^2-M^2}\gamma_{\mu_2}\frac{\slashed{k}+M}{k^2-M^2}\gamma_{\nu_1}\frac{\gamma_{\nu_3}}{(k^2-M^2)^2}\gamma_{\nu_2}\frac{\slashed{k}-\slashed{p}_1+M}{(k-p_1)^2-M^2}\right]\\
	&+8k_{\nu_3}k_{\mu_4}\tr\left[\gamma_{\mu_1}\frac{\slashed{k}+M}{k^2-M^2}\gamma_{\mu_2}\frac{\slashed{k}+M}{k^2-M^2}\gamma_{\nu_1}\frac{\slashed{k}+M}{(k^2-M^2)^3}\gamma_{\nu_2}\frac{\slashed{k}-\slashed{p}_1+M}{(k-p_1)^2-M^2}\right]\\	
	&-2k_{\nu_3}\tr\left[\gamma_{\mu_1}\frac{\slashed{k}+M}{k^2-M^2}\gamma_{\mu_2}\frac{\slashed{k}+M}{k^2-M^2}\gamma_{\nu_1}\frac{\gamma_{\mu_4}}{(k^2-M^2)^2}\gamma_{\nu_2}\frac{\slashed{k}-\slashed{p}_1+M}{(k-p_1)^2-M^2}\right]\\
	&-2k_{\mu_4}\tr\left[\gamma_{\mu_1}\frac{\slashed{k}+M}{k^2-M^2}\gamma_{\mu_2}\frac{\slashed{k}+M}{(k^2-M^2)^2}\gamma_{\nu_1}\frac{\gamma_{\nu_3}}{k^2-M^2}\gamma_{\nu_2}\frac{\slashed{k}-\slashed{p}_1+M}{(k-p_1)^2-M^2}\right]\\
	&+4k_{\nu_3}k_{\mu_4}\tr\left[\gamma_{\mu_1}\frac{\slashed{k}+M}{k^2-M^2}\gamma_{\mu_2}\frac{\slashed{k}+M}{(k^2-M^2)^2}\gamma_{\nu_1}\frac{\slashed{k}+M}{(k^2-M^2)^2}\gamma_{\nu_2}\frac{\slashed{k}-\slashed{p}_1+M}{(k-p_1)^2-M^2}\right]\\	
	&=\tr\left[\gamma_{\mu_1}\frac{\slashed{k}+M}{k^2-M^2}\gamma_{\mu_2}\frac{\gamma_{\mu_4}}{k^2-M^2}\gamma_{\nu_1}\frac{\gamma_{\nu_3}}{k^2-M^2}\gamma_{\nu_2}\frac{\slashed{k}-\slashed{p}_1+M}{(k-p_1)^2-M^2}\right]\\
	&-4k_{\nu_3}\tr\left[\gamma_{\mu_1}\frac{\slashed{k}+M}{k^2-M^2}\gamma_{\mu_2}\frac{\eta_{\mu_4\nu_1}(\slashed{k}+M)+k_{\nu_1}\gamma_{\mu_4}-\gamma_{\nu_1}k_{\mu_4}}{(k^2-M^2)^3}\gamma_{\nu_2}\frac{\slashed{k}-\slashed{p}_1+M}{(k-p_1)^2-M^2}\right]\\
	&-2\eta_{\nu_3\mu_4}\tr\left[\gamma_{\mu_1}\frac{\slashed{k}+M}{k^2-M^2}\gamma_{\mu_2}\frac{\slashed{k}+M}{k^2-M^2}\gamma_{\nu_1}\frac{\slashed{k}+M}{(k^2-M^2)^2}\gamma_{\nu_2}\frac{\slashed{k}-\slashed{p}_1+M}{(k-p_1)^2-M^2}\right]\\
	&-4k_{\mu_4}\tr\left[\gamma_{\mu_1}\frac{\slashed{k}+M}{k^2-M^2}\gamma_{\mu_2}\frac{\slashed{k}+M}{k^2-M^2}\gamma_{\nu_1}\frac{\gamma_{\nu_3}}{(k^2-M^2)^2}\gamma_{\nu_2}\frac{\slashed{k}-\slashed{p}_1+M}{(k-p_1)^2-M^2}\right]\\
	&+12k_{\nu_3}k_{\mu_4}\tr\left[\gamma_{\mu_1}\frac{\slashed{k}+M}{k^2-M^2}\gamma_{\mu_2}\frac{\slashed{k}+M}{k^2-M^2}\gamma_{\nu_1}\frac{\slashed{k}+M}{(k^2-M^2)^3}\gamma_{\nu_2}\frac{\slashed{k}-\slashed{p}_1+M}{(k-p_1)^2-M^2}\right]}
The $p_1$ appears seperately and we can compute
\eq{&\frac{\partial^2}{\partial p_1^{\mu_3}p_1^{\nu_4}}\frac{\slashed{k}-\slashed{p}_1+M}{(k-p_1)^2-M^2}=-\frac{\partial}{\partial p_1^{\mu_3}}\left[\frac{\gamma_{\nu_4}}{(k-p_1)^2-M^2}-2k_{\nu_4}\frac{\slashed{k}-\slashed{p}_1+M}{((k-p_1)^2-M^2)^2}\right]\\
	&=-\frac{2k_{\nu_4}\gamma_{\mu_3}+2k_{\mu_3}\gamma_{\nu_4}}{(k^2-M^2)^2}+8k_{\mu_3}k_{\nu_4}\frac{\slashed{k}+M}{(k^2-M^2)^3}\ .}
We get many terms but some of them will vanish due to antisymmetrization over the $\mu$s and the $\nu$s. We get
\eq{&\tr\left[\gamma_{\mu_1}\frac{\slashed{k}+M}{k^2-M^2}\gamma_{\mu_2}\frac{\gamma_{\mu_4}}{k^2-M^2}\gamma_{\nu_1}\frac{\gamma_{\nu_3}}{k^2-M^2}\gamma_{\nu_2}\left(-\frac{2k_{\nu_4}\gamma_{\mu_3}+2k_{\mu_3}\gamma_{\nu_4}}{(k^2-M^2)^2}+8k_{\mu_3}k_{\nu_4}\frac{\slashed{k}+M}{(k^2-M^2)^3}\right)\right]\\
	&+8\eta_{\mu_4\nu_1}k_{\mu_3}k_{\nu_3}\tr\left[\gamma_{\mu_1}\frac{\slashed{k}+M}{k^2-M^2}\gamma_{\mu_2}\frac{(\slashed{k}+M)}{(k^2-M^2)^3}\gamma_{\nu_2}\frac{\gamma_{\nu_4}}{(k^2-M^2)^2}\right]\\
	&-2\eta_{\nu_3\mu_4}\tr\left[\gamma_{\mu_1}\frac{\slashed{k}+M}{k^2-M^2}\gamma_{\mu_2}\frac{\slashed{k}+M}{k^2-M^2}\gamma_{\nu_1}\frac{\slashed{k}+M}{(k^2-M^2)^2}\gamma_{\nu_2}\left(-\frac{2k_{\nu_4}\gamma_{\mu_3}+2k_{\mu_3}\gamma_{\nu_4}}{(k^2-M^2)^2}+8k_{\mu_3}k_{\nu_4}\frac{\slashed{k}+M}{(k^2-M^2)^3}\right)\right]\\
	&+8k_{\nu_4}k_{\mu_4}\tr\left[\gamma_{\mu_1}\frac{\slashed{k}+M}{k^2-M^2}\gamma_{\mu_2}\frac{\slashed{k}+M}{k^2-M^2}\gamma_{\nu_1}\frac{\gamma_{\nu_3}}{(k^2-M^2)^2}\gamma_{\nu_2}\frac{\gamma_{\mu_3}}{(k^2-M^2)^2}\right]\\
	&=-\frac{8k_{\mu_3}k_{\nu_4}}{(k^2-M^2)^5}\tr\left[\gamma_{\mu_1}\gamma_{\mu_2}\gamma_{\mu_4}\gamma_{\nu_1}\gamma_{\nu_3}\gamma_{\nu_2}\right]-\frac{8\eta_{\mu_4\nu_1}k_{\mu_3}k_{\nu_3}}{(k^2-M^2)^5}\tr\left[\gamma_{\mu_1}\gamma_{\mu_2}\gamma_{\nu_2}\gamma_{\nu_4}\right]-\frac{16\eta_{\nu_3\mu_4}k_{\mu_3}k_{\nu_4}}{(k^2-M^2)^5}\tr\left[\gamma_{\mu_1}\gamma_{\mu_2}\gamma_{\nu_1}\gamma_{\nu_2}\right]\\
	&-\frac{2}{(k^2-M^2)^5}\tr\left[\gamma_{\mu_1}\slashed{k}\gamma_{\mu_2}\gamma_{\mu_4}\gamma_{\nu_1}\gamma_{\nu_3}\gamma_{\nu_2}(k_{\nu_4}\gamma_{\mu_3}+k_{\mu_3}\gamma_{\nu_4})\right]-\frac{8k_{\nu_4}k_{\mu_4}}{(k^2-M^2)^5}\tr\left[\gamma_{\mu_1}\gamma_{\mu_2}\gamma_{\nu_1}\gamma_{\nu_3}\gamma_{\nu_2}\gamma_{\mu_3}\right]\\
&-\frac{4\eta_{\nu_3\mu_4}k_{\nu_4}}{(k^2-M^2)^5}\tr\left[\gamma_{\mu_1}\slashed{k}\gamma_{\mu_2}\gamma_{\nu_1}\gamma_{\nu_2}\gamma_{\mu_3}\right]-\frac{4\eta_{\nu_3\mu_4}k_{\mu_3}}{(k^2-M^2)^5}\tr\left[\gamma_{\mu_1}\gamma_{\mu_2}\gamma_{\nu_1}\slashed{k}\gamma_{\nu_2}\gamma_{\nu_4}\right]\ .}
At this point it will be useful to perform the integration over $k$ using the formula
\eq{\int\frac{d^4k}{(2\pi)^4}\frac{k_\mu k_\nu}{(k^2-M^2)^n}=\frac{(-1)^{n-1}i}{(4\pi)^{d/2}}\frac{\eta_{\mu\nu}}{2}\frac{\Gamma(n-d/2-1)}{\Gamma(n)}\onov{M^{2n-d-2}}}
and get
\eq{&-\frac{i\eta_{\mu_3\nu_4}}{6(4\pi)^2M^4}\tr\left[\gamma_{\mu_1}\gamma_{\mu_2}\gamma_{\mu_4}\gamma_{\nu_1}\gamma_{\nu_3}\gamma_{\nu_2}\right]-\frac{i\eta_{\mu_4\nu_1}\eta_{\nu_3\mu_3}}{6(4\pi)^2M^4}\tr\left[\gamma_{\mu_1}\gamma_{\mu_2}\gamma_{\nu_2}\gamma_{\nu_4}\right]-\frac{i\eta_{\nu_3\mu_4}\eta_{\mu_3\nu_4}}{3(4\pi)^2M^4}\tr\left[\gamma_{\mu_1}\gamma_{\mu_2}\gamma_{\nu_1}\gamma_{\nu_2}\right]\\
	&-\frac{i}{24(4\pi)^2M^4}\tr\left[\gamma_{\mu_1}\gamma_{\nu_4}\gamma_{\mu_2}\gamma_{\mu_4}\gamma_{\nu_1}\gamma_{\nu_3}\gamma_{\nu_2}\gamma_{\mu_3}\right]-\frac{i}{24(4\pi)^2M^4}\tr\left[\gamma_{\mu_1}\gamma_{\mu_3}\gamma_{\mu_2}\gamma_{\mu_4}\gamma_{\nu_1}\gamma_{\nu_3}\gamma_{\nu_2}\gamma_{\nu_4}\right]\\&-\frac{i\eta_{\nu_4\mu_4}}{6(4\pi)^2M^4}\tr\left[\gamma_{\mu_1}\gamma_{\mu_2}\gamma_{\nu_1}\gamma_{\nu_3}\gamma_{\nu_2}\gamma_{\mu_3}\right]-\frac{i\eta_{\nu_3\mu_4}}{12(4\pi)^2M^4}\tr\left[\gamma_{\mu_1}\gamma_{\nu_4}\gamma_{\mu_2}\gamma_{\nu_1}\gamma_{\nu_2}\gamma_{\mu_3}+\gamma_{\mu_1}\gamma_{\mu_2}\gamma_{\nu_1}\gamma_{\mu_3}\gamma_{\nu_2}\gamma_{\nu_4}\right]\ .}
Now we will perform the traces and contract them with $\epsilon_{\mu_1\mu_2\mu_3\mu_4}\epsilon_{\nu_1\nu_2\nu_3\nu_4}$. We get
\eq{&\frac{6i}{\pi^2M^4}-\frac{6i}{\pi^2M^4}=0\ .}
On exactly the same way, also the second diagram in \ref{Feynman} with the order $A_1A_2B_2B_1$ contributes zero. The only diagram that may not vanish is the third diagram in \ref{Feynman} with the order $A_1B_1A_2B_2$. This diagram contributes
\eq{&P_{\mu_1\mu_2\nu_1\nu_2}^{a_1a_2b_1b_2}=-g_1^2g_2^2\tr(T^{a_1}T^{a_2})\tr(T^{b_1}T^{b_2})\int\frac{d^4k}{(2\pi)^4}\\
		&\tr\left[\gamma_{\mu_1}\frac{\slashed{k}+M}{k^2-M^2}\gamma_{\nu_1}\frac{\slashed{k}+\slashed{q}_1+M}{(k+q_1)^2-M^2}\gamma_{\mu_2}\frac{\slashed{k}+\slashed{p}_2+\slashed{q}_1+M}{(k+p_2+q_1)^2-M^2}\gamma_{\nu_2}\frac{\slashed{k}-\slashed{p}_1+M}{(k-p_1)^2-M^2}\right]\ .}
Again we will start by taking derivatives with respect to the momenta. We want
\eq{-\epsilon_{\mu_1\mu_2\mu_3\mu_4}\epsilon_{\nu_1\nu_2\nu_3\nu_4}\frac{\partial^4P}{\partial p_1^{\mu_3}\partial p_2^{\mu_4}\partial q_1^{\nu_3}\partial p_1^{\nu_4}}\ .}
Ignore the $g_1^2g_2^2\tr(T^{a_1}T^{a_2})\tr(T^{b_1}T^{b_2})\int\frac{d^4k}{(2\pi)^4}$ and act on the integrand. 
Start with $p_1$. The relevant computation is
\eq{&\frac{\partial^2}{\partial p_1^{\mu_3}\partial p_1^{\nu_4}}\frac{\slashed{k}-\slashed{p}_1+M}{(k-p_1)^2-M^2}=\frac{\partial}{\partial p_1^{\mu_3}}\left[\frac{-\gamma_{\nu_4}}{(k-p_1)^2-M^2}+2k_{\nu_4}\frac{\slashed{k}-\slashed{p}_1+M}{((k-p_1)^2-M^2)^2}\right]\\
&=-\frac{2k_{\mu_3}\gamma_{\nu_4}}{((k-p_1)^2-M^2)^2}-\frac{2k_{\nu_4}\gamma_{\mu_3}}{((k-p_1)^2-M^2)^2}+8k_{\mu_3}k_{\nu_4}\frac{\slashed{k}-\slashed{p}_1+M}{((k-p_1)^2-M^2)^3}\ .}
Taking $p_1=0$ we get
\eq{-\frac{2(k_{\nu_4}\gamma_{\mu_3}+k_{\mu_3}\gamma_{\nu_4})}{(k^2-M^2)^2}+8k_{\mu_3}k_{\nu_4}\frac{\slashed{k}+M}{(k^2-M^2)^3}\ .}
We will also compute the $p_2$ derivative:
\eq{\frac{\partial}{\partial p_2^{\mu_4}}\frac{\slashed{k}+\slashed{p}_2+\slashed{q}_1+M}{(k+p_2+q_1)^2-M^2}|_{p_2=0}=\frac{\gamma_{\mu_4}}{(k+q_1)^2-M^2}-2(k+q_1)_{\mu_4}\frac{\slashed{k}+\slashed{q}_1+M}{((k+q_1)^2-M^2)^2}\ .}
Now we need to compute the $q_1$ derivative. The relevant part is
\eq{\frac{\partial}{\partial q_1^{\nu_3}}\left[\frac{\slashed{k}+\slashed{q}_1+M}{((k+q_1)^2-M^2)^2}\gamma_{\mu_2}\gamma_{\mu_4}-2(k+q_1)_{\mu_4}\frac{\slashed{k}+\slashed{q}_1+M}{((k+q_1)^2-M^2)^3}\gamma_{\mu_2}(\slashed{k}+\slashed{q}_1+M)\right]\ .}
First, lets rewrite \eq{&(\slashed{k}+\slashed{q}_1+M)\gamma_{\mu_2}(\slashed{k}+\slashed{q}_1+M)=\gamma_{\mu_2}(-\slashed{k}-\slashed{q}_1+M)(\slashed{k}+\slashed{q}_1+M)+2(k+q_1)_{\mu_2}(\slashed{k}+\slashed{q}_1+M)\\
&=\gamma_{\mu_2}(M^2-(k+q_1)^2)+2(k+q_1)_{\mu_2}(\slashed{k}+\slashed{q}_1+M)\ .}
Plug it back in and get
\eq{\frac{\partial}{\partial q_1^{\nu_3}}\left[\frac{\slashed{k}+\slashed{q}_1+M}{((k+q_1)^2-M^2)^2}\gamma_{\mu_2}\gamma_{\mu_4}+\frac{2(k+q_1)_{\mu_4}\gamma_{\mu_2}}{((k+q_1)^2-M^2)^2}-4(k+q_1)_{\mu_4}(k+q_1)_{\mu_2}\frac{(\slashed{k}+\slashed{q}_1+M)}{((k+q_1)^2-M^2)^3}\right]\ .}
The last term will vanish after contracting with the $\epsilon$. We are left with
\eq{&\frac{\partial}{\partial q_1^{\nu_3}}\left[\frac{\slashed{k}+\slashed{q}_1+M}{((k+q_1)^2-M^2)^2}\gamma_{\mu_2}\gamma_{\mu_4}+\frac{2(k+q_1)_{\mu_4}\gamma_{\mu_2}}{((k+q_1)^2-M^2)^2}\right]\\
&=\frac{\gamma_{\nu_3}\gamma_{\mu_2}\gamma_{\mu_4}}{(k^2-M^2)^2}-\frac{4k_{\nu_3}(\slashed{k}+M)\gamma_{\mu_2}\gamma_{\mu_4}}{(k^2-M^2)^3}+\frac{2\eta_{\nu_3\mu_4}\gamma_{\mu_2}}{(k^2-M^2)^2}-\frac{8k_{\nu_3}k_{\mu_4}\gamma_{\mu_2}}{(k^2-M^2)^3}\ .}

Now plug it back into the integral and get
\eq{&\int\frac{d^4k}{(2\pi)^4}\tr\left[\gamma_{\mu_1}\frac{\slashed{k}+M}{k^2-M^2}\gamma_{\nu_1}\left(\frac{\gamma_{\nu_3}\gamma_{\mu_2}\gamma_{\mu_4}}{(k^2-M^2)^2}-\frac{4k_{\nu_3}(\slashed{k}+M)\gamma_{\mu_2}\gamma_{\mu_4}}{(k^2-M^2)^3}+\frac{2\eta_{\nu_3\mu_4}\gamma_{\mu_2}}{(k^2-M^2)^2}-\frac{8k_{\nu_3}k_{\mu_4}\gamma_{\mu_2}}{(k^2-M^2)^3}\right)\right.\\&\left.\gamma_{\nu_2}\left(-\frac{2(k_{\nu_4}\gamma_{\mu_3}+k_{\mu_3}\gamma_{\nu_4})}{(k^2-M^2)^2}+8k_{\mu_3}k_{\nu_4}\frac{\slashed{k}+M}{(k^2-M^2)^3}\right)\right]\ .}
We will open the brackets and keep only terms that don't vanish after contracting with the $\epsilon$: The integrand is
\eq{&\onov{(k^2-M^2)^6}8k_{\mu_3}k_{\nu_4}\gamma_{\mu_1}(\slashed{k}+M)\gamma_{\nu_1}\gamma_{\nu_3}\gamma_{\mu_2}\gamma_{\mu_4}\gamma_{\nu_2}(\slashed{k}+M)\\&-\onov{(k^2-M^2)^5}2\gamma_{\mu_1}(\slashed{k}+M)\gamma_{\nu_1}\gamma_{\nu_3}\gamma_{\mu_2}\gamma_{\mu_4}\gamma_{\nu_2}(k_{\nu_4}\gamma_{\mu_3}+k_{\mu_3}\gamma_{\nu_4})\\&+\onov{(k^2-M^2)^6}8k_{\mu_3}k_{\nu_3}\gamma_{\mu_1}(\slashed{k}+M)\gamma_{\nu_1}(\slashed{k}+M)\gamma_{\mu_2}\gamma_{\mu_4}\gamma_{\nu_2}\gamma_{\nu_4}\\&+\onov{(k^2-M^2)^6}16\eta_{\nu_3\mu_4}k_{\mu_3}k_{\nu_4}\gamma_{\mu_1}(\slashed{k}+M)\gamma_{\nu_1}\gamma_{\mu_2}\gamma_{\nu_2}(\slashed{k}+M)\\&-\onov{(k^2-M^2)^5}4\eta_{\nu_3\mu_4}\gamma_{\mu_1}(\slashed{k}+M)\gamma_{\nu_1}\gamma_{\mu_2}\gamma_{\nu_2}(k_{\nu_4}\gamma_{\mu_3}+k_{\mu_3}\gamma_{\nu_4})\ .}
We can do the following simplifications. First, only even number of $\gamma$ matrices will survive the trace, therefore in the second and last lines we can replace $(\slashed{k}+M)\to\slashed{k}$. The second simplification is to use
\eq{&(\slashed{k}+M)\gamma_{\mu}(\slashed{k}+M)=\gamma_{\mu}(M^2-k^2)+2k_{\mu}(\slashed{k}+M)\ ,} together with the cyclicity of the trace and the antisymmetrization over $\mu$ indices and over the $\nu$ indices. We get
\eq{&-\onov{(k^2-M^2)^5}8k_{\mu_3}k_{\nu_4}\gamma_{\mu_1}\gamma_{\nu_1}\gamma_{\nu_3}\gamma_{\mu_2}\gamma_{\mu_4}\gamma_{\nu_2}\\&-\onov{(k^2-M^2)^5}2\gamma_{\mu_1}\slashed{k}\gamma_{\nu_1}\gamma_{\nu_3}\gamma_{\mu_2}\gamma_{\mu_4}\gamma_{\nu_2}(k_{\nu_4}\gamma_{\mu_3}+k_{\mu_3}\gamma_{\nu_4})\\&-\onov{(k^2-M^2)^5}8k_{\mu_3}k_{\nu_3}\gamma_{\mu_1}\gamma_{\nu_1}\gamma_{\mu_2}\gamma_{\mu_4}\gamma_{\nu_2}\gamma_{\nu_4}\\&-\onov{(k^2-M^2)^5}16\eta_{\nu_3\mu_4}k_{\mu_3}k_{\nu_4}\gamma_{\mu_1}\gamma_{\nu_1}\gamma_{\mu_2}\gamma_{\nu_2}\\&-\onov{(k^2-M^2)^5}4\eta_{\nu_3\mu_4}\gamma_{\mu_1}\slashed{k}\gamma_{\nu_1}\gamma_{\mu_2}\gamma_{\nu_2}(k_{\nu_4}\gamma_{\mu_3}+k_{\mu_3}\gamma_{\nu_4})\ .}
At this point it is useful to perform the $k$ integral using the formula
\eq{\int\frac{d^4k}{(2\pi)^4}\frac{k_\mu k_\nu}{(k^2-M^2)^5}=\frac{i\eta_{\mu\nu}}{48(4\pi)^{2}M^{4}}\ .}
We get
\eq{&-\frac{i\eta_{\mu_3\nu_4}}{6(4\pi)^2M^4}\gamma_{\mu_1}\gamma_{\nu_1}\gamma_{\nu_3}\gamma_{\mu_2}\gamma_{\mu_4}\gamma_{\nu_2}\\&-\frac{i}{24(4\pi)^{2}M^{4}}\left[\gamma_{\mu_1}\gamma_{\nu_4}\gamma_{\nu_1}\gamma_{\nu_3}\gamma_{\mu_2}\gamma_{\mu_4}\gamma_{\nu_2}\gamma_{\mu_3}+\gamma_{\mu_1}\gamma_{\mu_3}\gamma_{\nu_1}\gamma_{\nu_3}\gamma_{\mu_2}\gamma_{\mu_4}\gamma_{\nu_2}\gamma_{\nu_4}\right]\\&-\frac{i\eta_{\mu_3\nu_3}}{6(4\pi)^{2}M^{4}}\gamma_{\mu_1}\gamma_{\nu_1}\gamma_{\mu_2}\gamma_{\mu_4}\gamma_{\nu_2}\gamma_{\nu_4}-\frac{i\eta_{\mu_3\nu_4}\eta_{\nu_3\mu_4}}{3(4\pi)^{2}M^{4}}\gamma_{\mu_1}\gamma_{\nu_1}\gamma_{\mu_2}\gamma_{\nu_2}\\&-\frac{i\eta_{\nu_3\mu_4}}{12(4\pi)^{2}M^{4}}\left[\gamma_{\mu_1}\gamma_{\nu_4}\gamma_{\nu_1}\gamma_{\mu_2}\gamma_{\nu_2}\gamma_{\mu_3}+\gamma_{\mu_1}\gamma_{\mu_3}\gamma_{\nu_1}\gamma_{\mu_2}\gamma_{\nu_2}\gamma_{\nu_4}\right]\ .}
Start from the terms with 8 $\gamma$. We can bring all the $\mu$s together by using \eq{\gamma_{\mu_2}\gamma_{\mu_4}\gamma_{\nu_2}=\gamma_{\nu_2}\gamma_{\mu_2}\gamma_{\mu_4}+2\gamma_{\mu_2}\eta_{\mu_4\nu_2}-2\gamma_{\mu_4}\eta_{\mu_2\nu_2}\to \gamma_{\nu_2}\gamma_{\mu_2}\gamma_{\mu_4}+4\gamma_{\mu_2}\eta_{\mu_4\nu_2}\ ,}after antisymmetrization. So we can write
\eq{&\gamma_{\mu_1}\gamma_{\nu_4}\gamma_{\nu_1}\gamma_{\nu_3}\gamma_{\mu_2}\gamma_{\mu_4}\gamma_{\nu_2}\gamma_{\mu_3}=\gamma_{\mu_1}\gamma_{\nu_4}\gamma_{\nu_1}\gamma_{\nu_3}\gamma_{\nu_2}\gamma_{\mu_2}\gamma_{\mu_4}\gamma_{\mu_3}+4\eta_{\mu_4\nu_2}\gamma_{\mu_1}\gamma_{\nu_4}\gamma_{\nu_1}\gamma_{\nu_3}\gamma_{\mu_2}\gamma_{\mu_3}\ ,\\
&\gamma_{\mu_1}\gamma_{\mu_3}\gamma_{\nu_1}\gamma_{\nu_3}\gamma_{\mu_2}\gamma_{\mu_4}\gamma_{\nu_2}\gamma_{\nu_4}=\gamma_{\mu_1}\gamma_{\mu_3}\gamma_{\nu_1}\gamma_{\nu_3}\gamma_{\nu_2}\gamma_{\nu_4}\gamma_{\mu_2}\gamma_{\mu_4}+4\eta_{\mu_4\nu_4}\gamma_{\mu_1}\gamma_{\mu_3}\gamma_{\nu_1}\gamma_{\nu_3}\gamma_{\nu_2}\gamma_{\mu_2}+4\eta_{\nu_2\mu_4}\gamma_{\mu_1}\gamma_{\mu_3}\gamma_{\nu_1}\gamma_{\nu_3}\gamma_{\mu_2}\gamma_{\nu_4}\ .}
Now we can use $\gamma_5=-\frac{i}{24}\epsilon^{\mu_1\mu_2\mu_3\mu_4}\gamma_{\mu_1}\gamma_{\mu_2}\gamma_{\mu_3}\gamma_{\mu_4}$ and $\tr(\gamma_5^2)=4$. After contractions with the $\epsilon$s and tracing, the 8$\gamma$ terms give together $\frac{12i}{\pi^2M^4}$. The term with 4 $\gamma$s gives 
\eq{&-\frac{i\eta_{\mu_3\nu_4}\eta_{\nu_3\mu_4}\epsilon^{\mu_1\mu_2\mu_3\mu_4}\epsilon^{\nu_1\nu_2\nu_3\nu_4}}{3(4\pi)^{2}M^{4}}\tr(\gamma_{\mu_1}\gamma_{\nu_1}\gamma_{\mu_2}\gamma_{\nu_2})=-\frac{2i(\eta^{\mu_1\nu_1}\eta^{\mu_2\nu_2}-\eta^{\mu_1\nu_2}\eta^{\mu_2\nu_1})}{3(4\pi)^{2}M^{4}}\tr(\gamma_{\mu_1}\gamma_{\nu_1}\gamma_{\mu_2}\gamma_{\nu_2})=0\ .}

We still need to compute all the terms with 6 $\gamma$s. 
\eq{&-\frac{i\eta_{\mu_3\nu_4}}{6(4\pi)^2M^4}\gamma_{\mu_1}\gamma_{\nu_1}\gamma_{\nu_3}\gamma_{\mu_2}\gamma_{\mu_4}\gamma_{\nu_2}\\&-\frac{i}{6(4\pi)^{2}M^{4}}\left[\eta_{\mu_4\nu_2}\gamma_{\mu_1}\gamma_{\nu_4}\gamma_{\nu_1}\gamma_{\nu_3}\gamma_{\mu_2}\gamma_{\mu_3}+\eta_{\mu_4\nu_4}\gamma_{\mu_1}\gamma_{\mu_3}\gamma_{\nu_1}\gamma_{\nu_3}\gamma_{\nu_2}\gamma_{\mu_2}+\eta_{\nu_2\mu_4}\gamma_{\mu_1}\gamma_{\mu_3}\gamma_{\nu_1}\gamma_{\nu_3}\gamma_{\mu_2}\gamma_{\nu_4}\right]\\&-\frac{i\eta_{\mu_3\nu_3}}{6(4\pi)^{2}M^{4}}\gamma_{\mu_1}\gamma_{\nu_1}\gamma_{\mu_2}\gamma_{\mu_4}\gamma_{\nu_2}\gamma_{\nu_4}-\frac{i\eta_{\nu_3\mu_4}}{12(4\pi)^{2}M^{4}}\left[\gamma_{\mu_1}\gamma_{\nu_4}\gamma_{\nu_1}\gamma_{\mu_2}\gamma_{\nu_2}\gamma_{\mu_3}+\gamma_{\mu_1}\gamma_{\mu_3}\gamma_{\nu_1}\gamma_{\mu_2}\gamma_{\nu_2}\gamma_{\nu_4}\right]\ .}
We will manipulate the terms using the antisymmetry of the indices and the cyclicity of the trace,
\eq{&-\frac{i\eta_{\mu_4\nu_4}}{4(4\pi)^2M^4}\gamma_{\mu_1}\gamma_{\nu_1}\gamma_{\nu_2}\gamma_{\mu_2}\gamma_{\mu_3}\gamma_{\nu_3}-\frac{i\eta_{\mu_4\nu_4}}{3(4\pi)^{2}M^{4}}\gamma_{\mu_1}\gamma_{\mu_2}\gamma_{\mu_3}\gamma_{\nu_1}\gamma_{\nu_2}\gamma_{\nu_3}+\frac{i\eta_{\nu_4\mu_4}}{12(4\pi)^{2}M^{4}}\gamma_{\mu_1}\gamma_{\mu_2}\gamma_{\nu_1}\gamma_{\nu_2}\gamma_{\mu_3}\gamma_{\nu_3}\\
&=\frac{i\eta_{\mu_4\nu_4}}{4(4\pi)^2M^4}\gamma_{\mu_1}\gamma_{\mu_2}\gamma_{\mu_3}\gamma_{\nu_1}\gamma_{\nu_2}\gamma_{\nu_3}-\frac{i\eta_{\mu_4\nu_4}}{3(4\pi)^{2}M^{4}}\gamma_{\mu_1}\gamma_{\mu_2}\gamma_{\mu_3}\gamma_{\nu_1}\gamma_{\nu_2}\gamma_{\nu_3}-\frac{i\eta_{\nu_4\mu_4}}{12(4\pi)^{2}M^{4}}\gamma_{\mu_1}\gamma_{\mu_2}\gamma_{\mu_3}\gamma_{\nu_1}\gamma_{\nu_2}\gamma_{\nu_3}\\
&-\frac{i\eta_{\mu_4\nu_4}\eta_{\mu_1\nu_3}}{2(4\pi)^2M^4}\gamma_{\nu_1}\gamma_{\nu_2}\gamma_{\mu_2}\gamma_{\mu_3}+\frac{i\eta_{\nu_4\mu_4}\eta_{\mu_3\nu_3}}{6(4\pi)^{2}M^{4}}\gamma_{\mu_1}\gamma_{\mu_2}\gamma_{\nu_1}\gamma_{\nu_2}\\
&=-\frac{i\eta_{\nu_4\mu_4}}{6(4\pi)^{2}M^{4}}\gamma_{\mu_1}\gamma_{\mu_2}\gamma_{\mu_3}\gamma_{\nu_1}\gamma_{\nu_2}\gamma_{\nu_3}-\frac{i\eta_{\mu_4\nu_4}\eta_{\mu_3\nu_3}}{3(4\pi)^2M^4}\gamma_{\mu_1}\gamma_{\mu_2}\gamma_{\nu_1}\gamma_{\nu_2}\ .}
Now taking the trace (and using the antisymmetry) we get
\eq{\frac{8i\eta_{\nu_4\mu_4}\eta_{\mu_1\nu_1}\eta_{\mu_2\nu_2}\eta_{\mu_3\nu_3}}{(4\pi)^{2}M^{4}}+\frac{8i\eta_{\mu_4\nu_4}\eta_{\mu_3\nu_3}\eta_{\mu_1\nu_1}\eta_{\mu_2\nu_2}}{3(4\pi)^2M^4}=\frac{2i\eta_{\mu_4\nu_4}\eta_{\mu_3\nu_3}\eta_{\mu_1\nu_1}\eta_{\mu_2\nu_2}}{3\pi^2M^4}\ .}
Contracting with the $\epsilon$s we get $-\frac{16i}{\pi^2M^4}$. Together with what we had before, it is $-\frac{4i}{\pi^2M^4}$.

To summarize the results so far, we found that
\eq{-\epsilon_{\mu_1\mu_2\mu_3\mu_4}\epsilon_{\nu_1\nu_2\nu_3\nu_4}\frac{\partial^4P_{\mu_1\mu_2\nu_1\nu_2}^{a_1a_2b_1b_2}}{\partial p_1^{\mu_3}\partial p_2^{\mu_4}\partial q_1^{\nu_3}\partial p_1^{\nu_4}}=-\frac{4i}{\pi^2M^4}g_1^2g_2^2\tr(T^{a_1}T^{a_2})\tr(T^{b_1}T^{b_2})\ ,}
where
\eq{P_{\mu_1\mu_2\nu_1\nu_2}^{a_1a_2b_1b_2}=\vev{A_{\mu_1}^{a_1}(p_1)A^{a_2}_{\mu_2}(p_2)B^{b_1}_{\nu_1}(q_1)B^{b_2}_{\nu_2}(-p_1-p_2-q_1)}\ .}
Comparing with
\eq{&-\epsilon_{\mu_1\mu_2\mu_3\mu_4}\epsilon_{\nu_1\nu_2\nu_3\nu_4}\frac{\partial^4P_{\mu_1\mu_2\nu_1\nu_2}^{a_1a_2b_1b_2}}{\partial p_1^{\mu_3}\partial p_2^{\mu_4}\partial q_1^{\nu_3}\partial p_1^{\nu_4}}=720iC\tr(T^{a_1}T^{a_2})\tr(T^{b_1}T^{b_2})\ ,}
 we find
\eq{C=-\frac{g_1^2g_2^2}{180\pi^2M^2}\ .}
We found that $C<0$, which means that in the groundstate, $F_A\wedge F_A$ and $F_B\wedge F_B$ have the same sign.

\bibliography{Bifundamentalbib}

	\end{document}